# Silicon sampling answers with country-level assumptions, not individual attitudes: Cross-national evidence from the European Social Survey

Chuyao Wang

Department of Methodology and Data Science Institute, London School of Economics and Political Science, London, United Kingdom

Correspondence: Chuyao Wang, c.wang85@lse.ac.uk.

## Abstract

Silicon sampling uses large language models (LLMs) to simulate survey respondents. Whether it recovers cross-national variation, and why, remains unresolved. This study evaluates it against European Social Survey Round 11 (30 countries, 42 items) with two open-weight LLMs under first- and third-person prompts, plus backstory and response-format experiments. Aggregate recovery is moderate and uneven across items. Adding the country name to a three-variable demographic backstory raises the median per-item correlation between simulated and observed country means from −0.03 to 0.52, and the richer profiles tested add no consistent gain. The respondent's country label acts as a country-level assumption that respondent detail does not revise. Naming the response-scale endpoints in words stops the model from ranking countries backwards, so the answer format sets the direction of the ranking. Individual-level recovery remains negligible in every condition and does not track aggregate recovery across countries. An average of neighboring countries, using no LLM, recovers country levels more accurately than every model condition and ranks them about as well. Silicon sampling can thus support exploratory country-ranking comparison after item-level validation and with the response format reported. It does not support individual or distributional inference.

**Keywords**: silicon sampling; large language models; cross-national variation; survey methodology

## 1 Introduction

Silicon sampling conditions a large language model (LLM) on a respondent's demographic profile so that it generates responses approximating those of a comparable human (Argyle et al., 2023). The approach is proposed as a cheaper and faster alternative to survey data collection, at a time of rising costs and falling response rates in cross-national surveys (Curtin et al., 2005; Jabkowski & Cichocki, 2025). The same capabilities work in reverse, as AI agents can now pass as human respondents and evade standard quality checks (Westwood, 2025).

Early evaluations reported that silicon sampling reproduced selected findings from human-subject studies (Aher et al., 2023). Argyle et al. (2023) report that GPT-3, when conditioned on sociodemographic backstories from the American National Election Studies, produces silicon samples whose correlational patterns mirror those of human respondents on a set of political items. Dillion et al. (2023) likewise document that model judgments of the wrongness of everyday actions track human ratings. In experimental settings, Ashokkumar et al. (2026) show that GPT-4 can predict treatment effects in 70 nationally representative United States survey experiments.

Later evaluations found compressed distributions and misrepresented groups. Bisbee et al. (2024) show that synthetic responses exhibit compressed variance, prompt sensitivity, and regression coefficients that diverge from human data. Dominguez-Olmedo et al. (2024) find strong ordering and labeling effects and near-uniform adjusted response distributions. Wang et al. (2025) document flattened within-group heterogeneity and misrepresentation of marginalized groups. Cross-nationally, model outputs remain closer to Western and English-speaking populations (Durmus et al., 2023; Tao et al., 2024).

Gaps in the existing silicon sampling literature limit what it can establish. First, most evaluations are centered on the United States (Argyle et al., 2023; Bisbee et al., 2024; Santurkar et al., 2023), and the main European validation covers one country and one outcome (von der Heyde et al., 2026). Second, many studies examine a narrow attitudinal domain, leaving item-level heterogeneity poorly characterized. Third, existing work shows that a change of country label moves model responses, but it neither isolates the contribution of the country label to recovery experimentally nor assesses subsequent profile enrichment across a broad survey battery (Qu & Wang, 2024). Fourth, individual-level recovery is usually reported as one figure pooled across respondents, leaving open whether it varies across countries and whether it aligns with aggregate recovery.

These gaps converge on two questions: what supplies the aggregate signal, and whether that signal reaches the individual respondents behind the country means. A country label may activate associations with the country named, or compositional differences across national samples may generate the cross-national variation on their own. This paper measures silicon sampling's recovery at both levels. It characterizes aggregate recovery across 30 countries and 42 items. A backstory conditioning experiment isolates the country label's contribution from demographic composition and from profile enrichment. A 2 × 2 design crosses the country label with the response format, turning the interpretation of the item inversions into a test. Estimating within-country individual recovery for each country then tests whether its geographic pattern matches the aggregate one.

This paper finds that the country label acts as a country-level assumption. The label produces most of the aggregate recovery, and enriching the profile beyond it adds no consistent gain. The response format decides whether the model ranks countries in the correct order or the reverse. Individual-level recovery remains negligible, and a regional average built from survey data alone recovers the same country means better.

## 2 Relevant literature and research questions

### 2.1 Evaluating silicon sampling across countries

Cross-national comparison is central to how survey data on attitudes are used. Political support, from trust in institutions to satisfaction with how democracy works, is compared across countries in accounts of democratic legitimacy (Norris, 2011), and political values in work on modernization and cultural change (Dalton & Welzel, 2014; Inglehart & Welzel, 2005). Subjective well-being likewise varies systematically with national economic, institutional, and cultural conditions (Diener et al., 1995). A comparative claim rests on how countries are ordered as much as on any single country's value, so a silicon sample must place countries in the right order and reproduce their levels.

Existing evaluations are narrow in the countries or the items they cover. Qu and Wang (2024) compare one environmental item across six World Values Survey countries, finding higher agreement in Western, English-speaking, and developed settings. Geng et al. (2024) simulate nine European Social Survey (ESS) Round 10 items for four countries and find that the simulated responses vary far less than the human ones, and Boelaert et al. (2025) cover five World Values Survey countries and four items. Cao et al. (2025) use two global surveys and find that specialization improves distributional correspondence, although performance remains weaker on unseen questions. Pataranutaporn et al. (2025) report the same Western and English-speaking advantage for individual-level predictions, though their evidence covers well-being alone. None of this work brings item-level ranking, individual-level recovery, and the contribution of country conditioning together on the same respondents.

Studies with wider country coverage show the same unevenness. Durmus et al. (2023) find that default model responses are closer to some national distributions than others, and that country prompting shifts responses without bringing them into line. Tao et al. (2024) compare five GPT models with values data from 107 countries and territories. Default outputs align most closely with English-speaking and Protestant European profiles, and cultural prompting improves correspondence for most countries without eliminating cross-country error.

Recovery should also be uneven across items, and part of that unevenness should follow how the response scale is presented, because option format shifts model answers independently of content (Section 2.2).

RQ1. How accurately does silicon sampling recover public attitudes at the aggregate level, in order and in level, and how far does that accuracy vary from item to item?

## 2.2 Candidate sources of aggregate recovery

LLMs exhibit cultural bias: their default outputs correspond more closely to the values and associations of some populations than of others. Internet-scale training data and model-development choices draw unevenly on languages and regions, making English-language and Western perspectives especially salient (Bender et al., 2021; Tao et al., 2024). Naous et al. (2024), for example, find that multilingual and Arabic models favor Western over Arab cultural entities even in Arabic-language tasks. Uneven correspondence also appears within countries, where Santurkar et al. (2023) find that human-feedback-tuned models correspond most closely to wealthier, educated, and politically liberal Americans.

The country label can move the output toward a country's measured attitudes or toward what is said about that country. Changing the country in the prompt shifts model responses (Qu & Wang, 2024; Tao et al., 2024). The two channels are observationally the same. The second channel is independently attested: human national-character ratings agree across raters while diverging from population mean traits (Terracciano et al., 2005), and LLMs given a nationality persona evaluate nations with a systematic valence, favoring Western Europe over Eastern Europe, Latin America, and Africa (Kamruzzaman & Kim, 2025). Agreement with national averages therefore does not distinguish the two.

Demographic conditioning works only if the model maps attributes onto the attitudes that go with them, so a richer profile should recover more. Argyle et al. (2023) remove backstory elements one at a time, but only for predicting United States vote choice. Adding partisanship and ideology lowers prediction error, while demographic attributes add little once politics is in the prompt (Bisbee et al., 2024). Models told to emulate different sociodemographic groups return nearly the same answer distribution whichever group they are given (Boelaert et al., 2025). A standard demographic profile has not been tested against the country name on a common cross-national battery.

The response format can create cross-national differences on its own. Format means how the answer scale is presented: numbers alone, or numbers whose endpoints are named in words. Reverse-coded items are a known source of misresponse in human surveys, where a minority of respondents answer them as though the direction had not changed (Weijters & Baumgartner, 2012). Language models show an analogous sensitivity to option format and position (Dominguez-Olmedo et al., 2024; Röttger et al., 2024; Tjuatja et al., 2024; Zheng et al., 2024). A numeric range without verbal endpoints leaves the scale direction to be inferred. Where numbers then carry different meanings for human and model, the comparison fails the measurement-equivalence condition (Davidov et al., 2014; King et al., 2004).

Aggregate recovery could therefore come from the composition of the national samples, from the country label, or from the response format. The backstory conditioning experiment separates the first two by testing whether the label adds recovery beyond composition. If the label activates calibrated knowledge that richer detail refines, recovery should rise with each level of enrichment. If it activates a country-level assumption, a single expectation attached to the country name, the label should produce one large step, and further detail should add little. That account also predicts that the label is specific to the country named: a wrong name should lower recovery for the respondents' real country. The 2 × 2 experiment tests a different account. If the numeric scale is read in the wrong direction on reverse-coded items, the same country knowledge will rank countries backwards on those items and correctly elsewhere. Supplying verbal endpoints should then remove the inversion without touching the model's knowledge.

RQ2. Which part of the prompt produces the aggregate signal, and why does the model rank countries backwards on some items?

### 2.3 Individual-level recovery and its relation to aggregate recovery

Individual-level recovery is more demanding than aggregate recovery, since model responses must preserve both the spread of human responses and the association between profile attributes and attitudes within populations. Bisbee et al. (2024) document compressed variance and unstable regression coefficients. Dominguez-Olmedo et al. (2024) show that response-format effects can dominate the resulting distributions. Wang et al. (2025) find that identity prompting flattens within-group heterogeneity and can reproduce out-group portrayals, not in-group responses.

Asking the model to be the person is the first-person (1P) framing, in which it speaks from the profile; asking it to predict the person is the third-person (3P) framing, in which it reasons about the profile. If individual recovery depends on mapping attributes to attitudes, the two framings could differ. A country-level assumption, by contrast, makes the framing irrelevant, so under that account the two should not differ.

Improved individual recovery may come from information a demographic profile does not carry. LLM agents grounded in interview or survey self-reports reproduce respondents' answers at 82–86% of two-week test-retest consistency and reduce racial and ideological accuracy gaps relative to demographics-only agents (Park et al., 2024). Fine-tuned models improve mainly by learning annotator-specific patterns, not demographic ones (Orlikowski et al., 2025), and sociodemographic prompting alone does not reliably improve subjective judgment (Sun et al., 2025). Whether sparse demographic conditioning recovers within-country differentiation, and how this varies across countries, remains open.

Whether individual recovery tracks aggregate recovery across countries depends on why they differ. If unequal country representation is the common constraint, countries with stronger aggregate recovery should also show stronger individual recovery (Durmus et al., 2023; Pataranutaporn et al., 2025; Qu & Wang, 2024). Aggregation alone predicts no such relation: idiosyncratic errors cancel in country means, so aggregate recovery can rest on group-level regularities that individual prediction cannot use. Structural inconsistencies across marginal, multivariate, and subgroup statistics point the same way (Li et al., 2025; Williams et al., 2026). The reviewed literature does not test that relation using the same respondents, items, and prompts.

RQ3. How far do the response distribution and individual-level recovery reproduce the human data? Does first- versus third-person framing change individual recovery, and does individual recovery track aggregate recovery across countries?

## 3 Data and methods

### 3.1 Survey benchmark

The benchmark is the European Social Survey, a cross-national probability-based survey (European Social Survey European Research Infrastructure, 2026). This paper uses data from ESS Round 11, which covers 30 countries. The analysis includes 42 survey items spanning 12 domains: social trust, institutional trust, political attitudes, political efficacy, subjective well-being, immigration attitudes, social values, national attachment, health, religion, safety, and income. The full item list with question text, scales, and domain classification is reported in Appendix Table A1. Figure 1 illustrates the simulation pipeline.

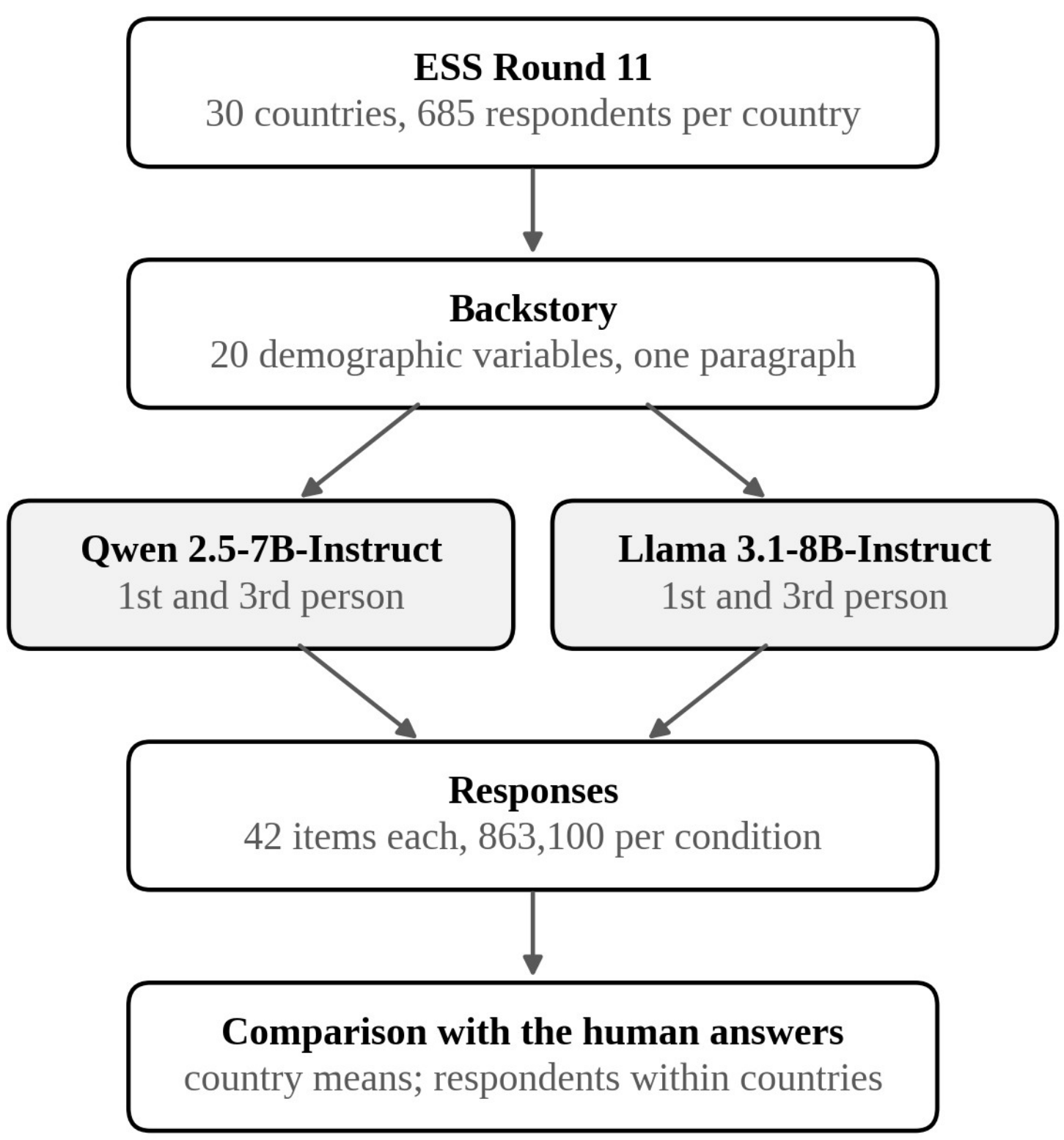

Figure 1: The simulation pipeline.

*Note:* ESS respondents are subsampled, assigned demographic backstories, and prompted through two models under two framing conditions. Model outputs are compared against human responses at the individual and country level.

### 3.2 Research design

The experiment is a 2 (model: Qwen, Llama) × 2 (prompt: first person, third person) crossing. Two open-weight instruction-tuned models are tested: Qwen 2.5-7B-Instruct (Alibaba) and Llama 3.1-8B-Instruct (Meta). Both are run locally on a university high-performance computing cluster (NVIDIA A100) using vLLM for batched inference. Temperature, which controls how random the output is, is set to 0.7, and respondent subsampling uses a fixed seed of 888, while generation itself is unseeded, except in the replicate arm (seed 889).

In the first-person (1P) condition, the model is instructed to adopt the respondent's persona and answer directly. In the third-person (3P) condition, the model is asked to predict how the described person would respond. Both prompts include the demographic backstory, the item text, the response scale, and a constraint to output a single number. The three social-trust items ask whether most people can be trusted, whether they try to be fair, and whether they are helpful. These three items carry the full ESS question wording, which embeds verbal endpoint descriptions. Every other item is presented with the abbreviated codebook wording reported in Appendix Table A1. The scale line for the other 39 items gives the numeric range as digits (e.g., “Scale: 0-10”) and names no endpoints in words.

For each country, 685 respondents are randomly subsampled from the ESS. That is the number available in the smallest country sample, so it is the largest subsample an equal-sized design can draw. The subsample is drawn with no design weights. The human benchmark is weighted by the ESS post-stratification weight (pspwght) throughout the main analyses, and the sensitivity of the results to that choice is reported in Section 4.4. Two random halves of the human sample agree at a median between-country correlation of 0.98, so the quantity being estimated is itself nearly noiseless. A single country-item cell has 80% power to detect a within-country correlation of 0.11, and pooling cells lowers that detectable correlation to 0.02, so the near-zero individual recovery in Section 4.3 is a measured near-zero.

Each respondent is prompted for all 42 items. Across the four model and prompt combinations the main experiment comprises 3.5 million response attempts, and 34.6 million across all 43 arms, each a separate generation run, including the backstory, swapped-label, and response-scale arms described below. Responses with both a valid human answer and a parseable model answer covered about 98% of those scheduled in every condition except Llama 3P, where coverage was 76%. The shortfall falls on the wide response scales, so that condition cannot identify a framing effect.

### 3.3 Backstory construction

Each respondent is described using a natural-language backstory generated from 20 ESS demographic variables: country, gender, age, birth year, marital status, domicile type, region, birth country, citizenship, education level, education years, main activity, household income decile, left-right political orientation, party affiliation, household size, children at home, trade union membership, discrimination group membership, and internet use frequency. These are standard sociodemographic measures commonly collected in probability surveys. An earlier 27-variable design included seven variables that overlapped with the outcome items; removing them left the 20 above (Appendix Table A2).

All backstories, prompts, and question texts are in English. Each backstory is constructed by mapping ESS codes to natural-language sentences with fixed value-label dictionaries (e.g., gender code 2 becomes “I am female”). Party affiliation is represented generically, with no country-specific party named, which may reduce individual-level recovery. The sentences form one paragraph, using the same fixed mapping across countries. The 3P prompt replaces the first line with “Consider a person with the following background:” and rephrases the question as “how would this person respond?”

Prompt language can shift cultural outputs, though not survey-prediction quality in one test (Boelaert et al., 2025; Lu et al., 2025). An abbreviated example prompt is provided in Appendix B. The replication materials provide the analysis-ready model summaries, the analysis and figure code, and the raw model responses paired with their human benchmark values, and are available at https://github.com/chuyao-wang/silicon-replication.

### 3.4 Backstory conditioning experiment

The backstory conditioning experiment asks whether aggregate recovery comes from the country label or from compositional differences across national samples. It uses Qwen 1P throughout, with the country label pair and the 2 × 2 replicated under Llama 1P (Section 4.2). The design is a set of paired contrasts on a common full profile, each observed with and without one block on the same respondents and items. One pair differs only by the country sentence, one only by the region code, and one only by the political identity block. Both arms of the country pair omit the region code, because the ESS region variable begins with the country code and would restore the country by another route.

The same with-and-without comparison runs over the whole 20-variable individual profile. These variables are grouped into 10 blocks: the base (gender, age, birth year), the country label, the region code, political identity, the socioeconomic block (education, main activity, and income), union membership with internet use, household composition (marital status, household size, and children at home), migration and citizenship, minority status, and domicile. The block is the unit of analysis because several variables are redundant with one another (age with birth year, education level with education years, household size with children at home); single variables are tested only where a block holds no such pair.

Each non-base block is tested at two positions: added alone to the base plus the country label (its earliest possible entry), and removed alone from the otherwise full profile (its latest possible exit). Together they bound the block’s contribution at intermediate positions only if that contribution changes monotonically as the profile fills in, which the design does not test.

Further arms test what the two positions alone cannot: a country-only arm drops the base entirely, asking whether the label needs any demographic context to work, and two arms state the same age, one in years and one as a birth year, asking whether a redundant variable's surface form matters.

A swapped-label arm replaces every backstory's country sentence with a wrong country, reassigning the 30 countries so that none keeps its own label, and leaves the rest of the profile intact. The arm is a placebo-treatment test of the composition account, which predicts that a wrong label makes no difference (Eggers et al., 2024), and it asks whether the model's sensitivity to the country label is specific to the country named. Nationality-assigned personas have been manipulated before (Kamruzzaman & Kim, 2025), and Qu and Wang (2024) relabeled United States respondents as Japanese and scored the result against the true respondents.

A 2 (country label: present, absent) × 2 (response scale: numbers only, numbers with verbal endpoints) experiment tests whether the format decides the direction of the label's effect. Its four arms run on a batch of 22 items, fixed in advance: 12 of the 13 reverse-coded items (vote has no scale direction), six forward-coded controls, and four items whose question wording already fixes the direction of the scale. Two reverse items whose model answers are near-constant are dropped from the primary contrast, leaving 10.

### 3.5 Evaluation metrics

Recovery denotes the correspondence between model-generated and human survey responses, at the level of aggregation a question concerns: country means for aggregate recovery, respondents within a country for individual recovery. It is assessed as linear association, distributional spread, and mean-level bias. Pearson correlations are used as a transparent summary of association, as in other aggregate comparisons (Ashokkumar et al., 2026; Dillion et al., 2023).

The primary aggregate measure, $r_{bc}$, is the per-item correlation across the 30 country means. The primary individual measure, $r_{wc}$, is the within-country correlation across respondents, computed per country-item cell. A pooled within-country correlation is taken across respondents from every country after centering within country and is used in Figure 6(b). A pooled correlation across all 1,260 country-item cells is reported as a labeled counterexample, because pooling lets scale differences between items be read as cross-national signal. The 30 countries and 42 items are the full set of interest, so the correlations themselves are reported without intervals. Intervals appear on contrasts: a 95% percentile bootstrap resampling items, or country-item cells where the contrast is defined on them, and a Fisher-z interval for correlations across the 30 countries. Appendix D defines one further measure used only there.

Two identical unseeded runs differ by a median absolute 0.060 in $r_{bc}$ per item, which sets the noise floor, or 0.079 on the Fisher-z scale, a transformation that makes differences between correlations comparable. Regressing each human response on five demographic predictors with country fixed effects gives a median within-country multiple correlation of 0.210, the benchmark for what demographic predictors recover linearly at the individual level. Two comparators that use no LLM, built from regional neighbors and from log GDP per capita, set the standard of adequacy (Section 4.1).

Distributional spread is evaluated by the ratio of model to human standard deviations for each country-item pair, and mean-level bias by the model-minus-human difference. A correlation ignores whether the model puts countries at the right level and spread, so Lin's concordance coefficient is also reported: it credits agreement only where the levels match as well as the ranking. Item-level aggregate recovery is read against $r_{bc}$ values of 0.5 and 0.7, and distributional recovery against a 0.9–1.1 band on the standard-deviation ratio. These thresholds were fixed before any counts were taken. Model and prompt contrasts use paired t-tests across the 42 items.

## 4 Results

### 4.1 Aggregate recovery is moderate and item-dependent

The aggregate recovery of silicon sampling is moderate (RQ1), and its size depends on the measure. Qwen with first-person prompting (1P) is the reference for the focused analyses below: Qwen recovers more than Llama, and the item-level pattern is similar across conditions. Under Qwen 1P, the pooled correlation is 0.822 while the median item-level correlation is 0.443. Pooling puts items of different scale lengths in one scatter, so the correlation it produces reflects differences between item means (Figure 2(a)). Against the thresholds of Section 3.5, 20 of 42 items exceed 0.5 and nine exceed 0.7.

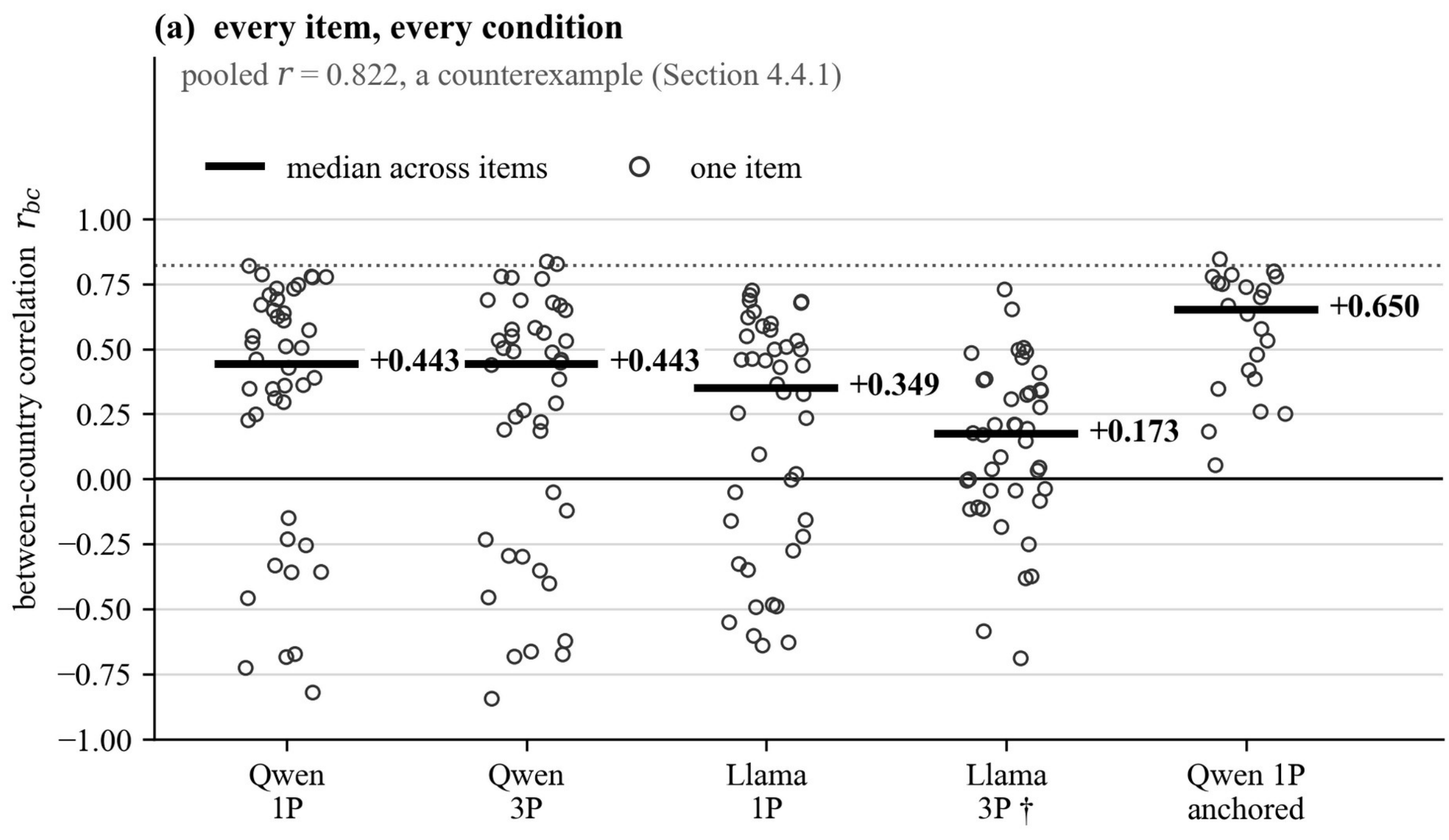

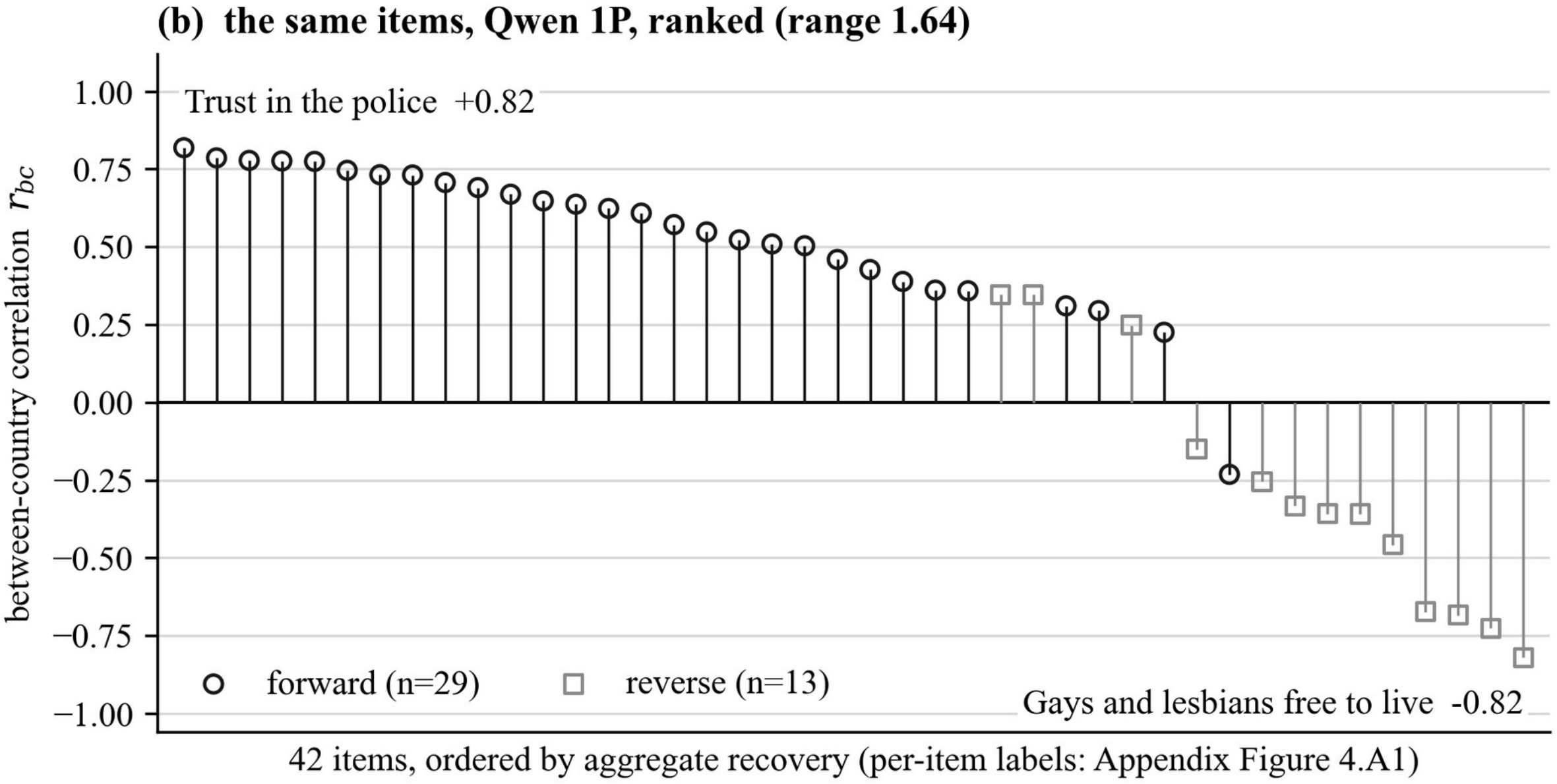

Figure 2: Aggregate recovery by condition and item by item.

*Note:* Panel (a) shows every item's between-country correlation in each condition, with the median across items marked. The pooled correlation is drawn once, as a labeled counterexample (all pooled $p < 0.001$). The Llama 3P arm (†) loses 23% of its scheduled responses to a parse failure and is not read substantively (Section 4.4). The anchored arm repeats Qwen 1P with the two scale endpoints named in words, on the 22-item batch of Section 3.4. Its level is not comparable with the 42-item arms. Panel (b) shows the same 42 items under Qwen 1P, ranked, with the highest and lowest named and reverse-coded items marked. Figure 6 is the individual-level counterpart, drawn on the scale of panel (b).

Recovery also varies across countries. For each country, the profile correlation is the correlation between its 42 human and model item means. The full ranking appears with the individual-level comparison in Section 4.3 (Figure 7). Under Qwen 1P, this correlation ranges from 0.69 in Bulgaria to 0.91 in Finland. The band is narrow, so the measure does not discriminate well enough to identify which countries the model recovers reliably.

Aggregate recovery is uneven across items: institutional trust and subjective well-being recover best, whereas income, safety, and health invert. The full labeled ranking is in Appendix Figure A1. Trust items dominate the top, with police, parliament, and the legal system all near 0.80. The model inverts the cross-national ordering of 11 items, and the two most severe are whether gay men and lesbians should be free to live as they wish (−0.82) and whether same-sex couples should have the right to adopt (−0.73).

Item direction accounts for most of these inversions. Of the 11 inversions, 10 are among the 13 reverse-coded items, whose numeric scale runs against the direction of the question, and direction alone explains 66% of the variance in $r_{bc}$ across the 42 items. The remaining one, whether European unification should go further (−0.23), is a bipolar item whose numeric direction is a convention, and in Figure 2(b) it is the only circle that is not reverse-coded in the negative tail. Among forward-coded items, recovery also rises with the size of the true cross-national differences (r = 0.46). Section 4.2 tests the effects of scale inversion experimentally.

The country ordering depends on the error metric. After rescaling each item to the unit interval, the normalized error, the mean absolute error on that scale, averages 0.19 across countries and ranges only from 0.16 to 0.22, and several Western European countries move into the higher-error group. Country ordering is nevertheless similar across Qwen and Llama ($r$ = 0.76 for normalized error, 0.97 for the profile correlation), so the same countries are easier or harder for both models even though the regional gradient is weak.

The model recovers levels and spread poorly even where rank correspondence is strongest. All 30 countries have positive mean bias under Qwen 1P, from 0.36 scale points in Finland to 1.40 in Bulgaria, with no clear geographic gradient. Lin's concordance coefficient shows that 87% of the agreement attainable at the observed rank correspondence is lost to bias and compression (Appendix F).

Predictors that use no LLM recover the same country means better. The average of a country's regional neighbors, excluding the country itself, gives a median $r_{bc}$ of 0.55 against the model's 0.44 and a median per-item normalized error of 0.056 against 0.180, winning on level accuracy for 41 of 42 items. Adding log GDP per capita to region dummies raises the comparator's median rbc to 0.62 and lowers its normalized error to 0.052. On ranking, the same 0.55 only narrowly exceeds the 0.52 of the country-only arm (Section 4.2), a gap inside the 0.060 noise floor (Section 4.4). The decisive comparison is therefore level accuracy. The model ranks countries with moderate accuracy, but estimates a country's value less accurately than the comparators do.

### 4.2 The country label produces aggregate recovery, and the response format sets its direction

A wrong country name lowers own-country recovery further than no country name does (RQ2). Scored against the respondents' own country, the swapped arm reaches a forward-item median $r_{bc}$ of 0.34, against 0.40 for the same profile with no country name at all. The median of the item-level differences is −0.096 [−0.142, −0.003], which is possible only if the model reads the label.

The wrong name is read without standing in for the country it names. Scored instead against the country it names, the same arm stays below the 0.56 that the true name reaches, and gains only +0.012 [−0.068, 0.122] over the own-country scoring. Under this reassignment the human means of the named and the true countries correlate at a median of −0.07, so one scoring does not follow from the other. Even under a wrong name, the remaining variables keep some of the respondents' own country in the profile. The no-country arm is not a zero-information control for the same reason (Appendix I; Figure 3).

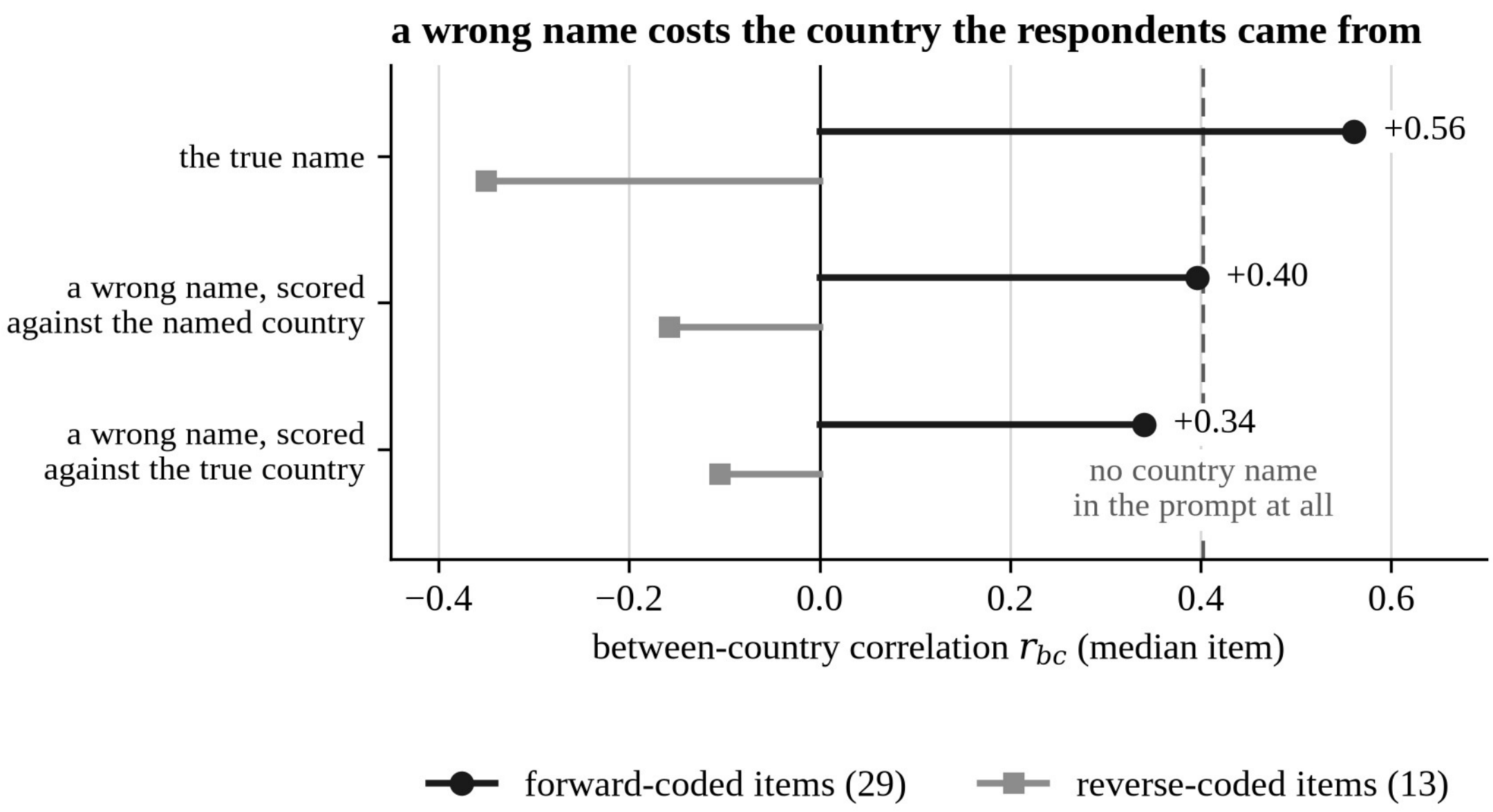

Figure 3: A wrong country name lowers own-country recovery, Qwen 1P.

*Note:* Median $r_{bc}$ over items, with forward- and reverse-coded items shown separately. The dashed line is the same profile with no country name in the prompt. The argument rests on the swapped arm scored against the respondents' own country, read against that line. Scoring the same arm against the country it names adds little, and the interval includes zero, so the arm is not evidence that recovery transfers to the country named.

Removing the country label moves recovery more than removing any other block does. With the country present the paired item-level correlation is higher by a median 0.135 on forward-coded items, where 27 of 29 improve ($p < 0.001$), and lower by 0.176 on reverse-coded items, where 1 of 13 improves. Of the three preset removal contrasts (country, region, and political identity), only the country label produces an effect that survives correction for multiplicity and exceeds the noise floor (Appendix H). The same removal replicates under Llama 1P, moving forward items by +0.25 and reverse items by −0.20 (Fisher z), with 25 of 29 forward items improving.

The country label alone carries almost the entire gain. Adding the country name to the base (gender, age, and birth year) raises median $r_{bc}$ from −0.03 to 0.52 (Fisher z +0.73 on forward items, −0.44 on reverse items), and nothing else approaches it at either position (Figure 4; Appendix Figure A3). The same 0.52 appears with no demographics at all, and stating the same age as a birth year leaves it unchanged, so the step depends neither on the base nor on how a variable is worded. The label's effect is also larger when it is added than when it is removed (+0.73 against +0.20, both Fisher z), because the full profile already carries some country information in the composition of its other variables (bounded in Section 5.4).

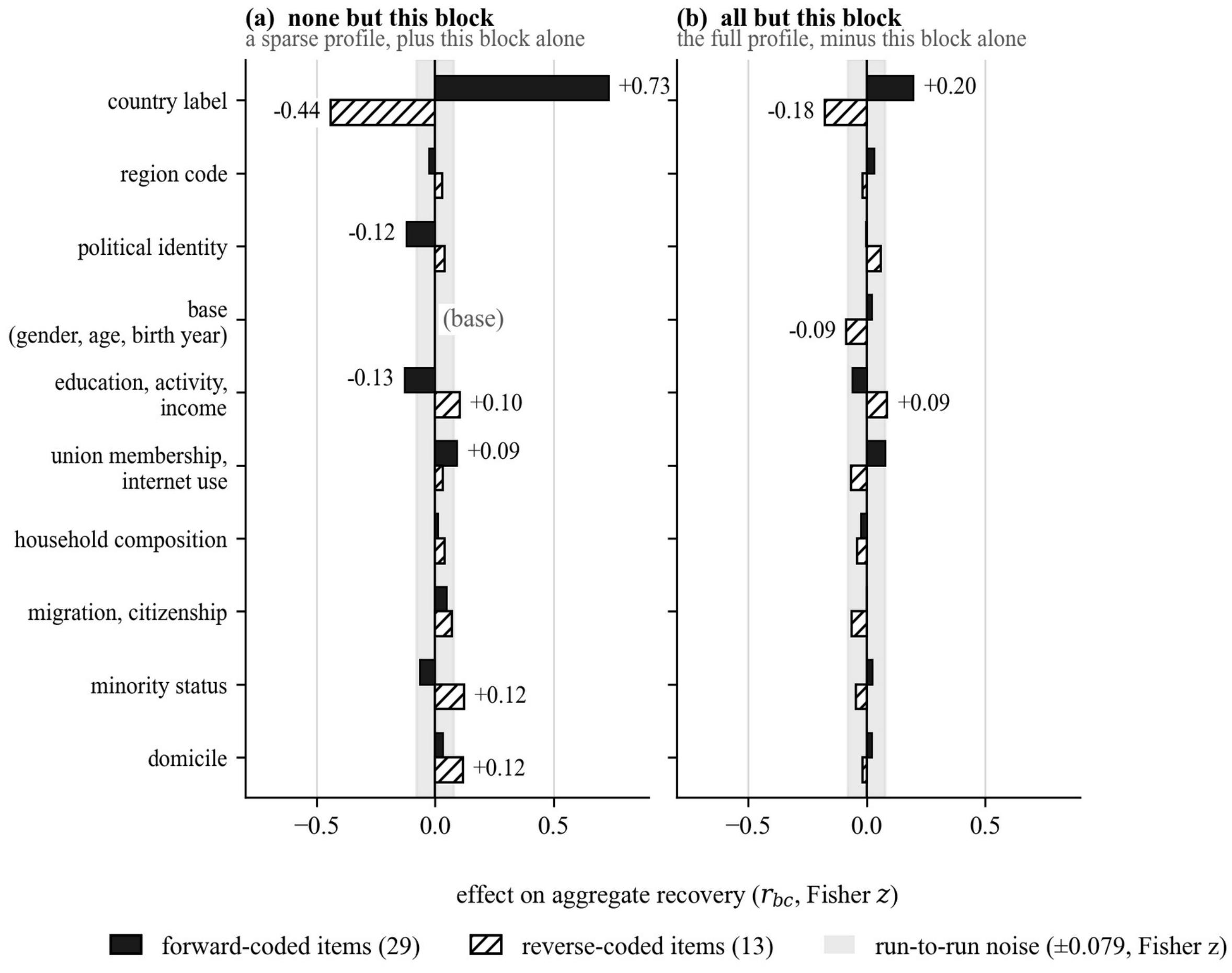

Figure 4: Each backstory block's effect on aggregate recovery, Qwen 1P.

*Note:* Each bar is a block's median effect on rbc (Fisher z) at one position. Panel (a) adds the block alone to the base plus the country label, and the country label to the base alone. Panel (b) removes it alone from the full profile. The two panels share one scale. The base has no bar in panel (a), because every other block is added to it. Reverse-coded items are drawn as separate bars because the effect reverses sign with wording. The shaded band is the Fisher-z noise floor (±0.079), and effects inside it are left unlabeled. The same contrast item by item is in Appendix Figure A3. Per-block values are provided with the figure data.

Adding political and socioeconomic information lowers aggregate recovery. Political identity lowers it when added to the base (−0.12) and has no detectable effect when removed from the full profile. The socioeconomic block lowers it at both positions (−0.13 added, −0.06 removed, the second inside the 0.079 Fisher-z noise floor), and removing household income alone reproduces nearly the whole block effect. Household income decile enters with that block, and income items recover worse with it than without, so in this design the added information lowers recovery even where it is directly relevant.

The full profile widens cross-national spread while leaving the correlation below the country-only level. The between-country dispersion ratio (model-to-human standard deviations across the 30 country means) rises from 0.22 with the country label alone to 0.28 with the full profile. Median within-country recovery nonetheless stays near zero in every arm, so the manipulation that produces the cross-national correspondence does not improve individual prediction.

The model may be reading the numeric scale in the wrong direction while holding the right country associations. That one error would produce both the inversions of Section 4.1 and the negative label effect above, and the 2 × 2 experiment tests it directly (Figure 5). Under numeric scales the country label helps forward-coded items and harms reverse-coded ones; under anchored scales (verbal endpoints) it helps both, improving every item that is not reverse-coded and 8 of the 10 that are. The triple difference (how much more the anchors change the label's effect on reverse items than on forward items) is +0.42 on the Fisher-z scale, with a 95% bootstrap interval of [0.13, 0.79]. The forward-reverse gap closes from +0.39 to −0.03 on the Fisher-z scale, which is inside the noise floor. Dropping the two largest movers leaves +0.16, and the same contrast in Llama 1P gives +0.45, both with intervals excluding zero (Appendix Table A5), so the result depends neither on two items nor on one model. The country label and the response format therefore interact.

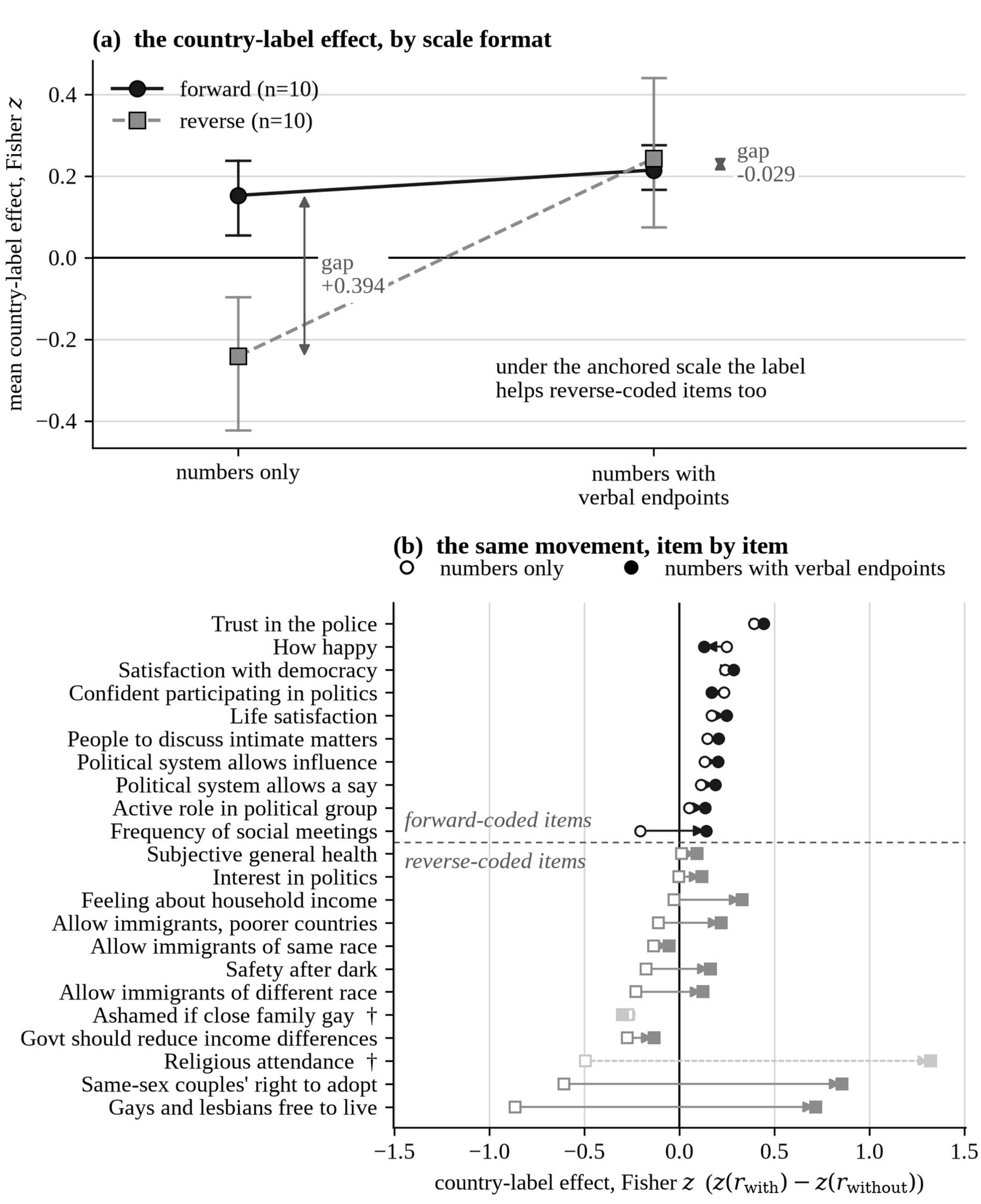

Figure 5: The country label by the response format, 22 items, Qwen 1P.

*Note:* A verbally anchored scale names the two endpoints of the response scale in words. The numeric arm gives the number range alone. Panel (a) shows the four cell means with 95% bootstrap intervals over items. Verbal anchors flip the label's effect on reverse-coded items from negative to positive. Panel (b) shows the same movement for every one of the 22 items in the batch, with the two prespecified exclusions marked (†). The triple difference excludes zero under all three prespecified reverse-item sets (Appendix Table A5).

### 4.3 Individual-level recovery is negligible and does not track aggregate recovery

Individual-level recovery remains near zero across model-prompt conditions (RQ3). Median within-country recovery does not exceed 0.028 (Qwen 1P) in any of the four model-prompt conditions, which implies less than 0.1% shared variance. With up to 685 respondents per country-item cell, many of these correlations are statistically significant yet substantively negligible. Demographic signal is present in the human data: five demographic predictors reach 0.210, well above what the model achieves (Figure 6). For Qwen, first- and third-person framing do not detectably differ: mean rwc is 0.027 against 0.029 ($p = 0.785$; Appendix Table A4). In a replicate arm, 95% of the within-country variance that the model produces is random variation between runs.

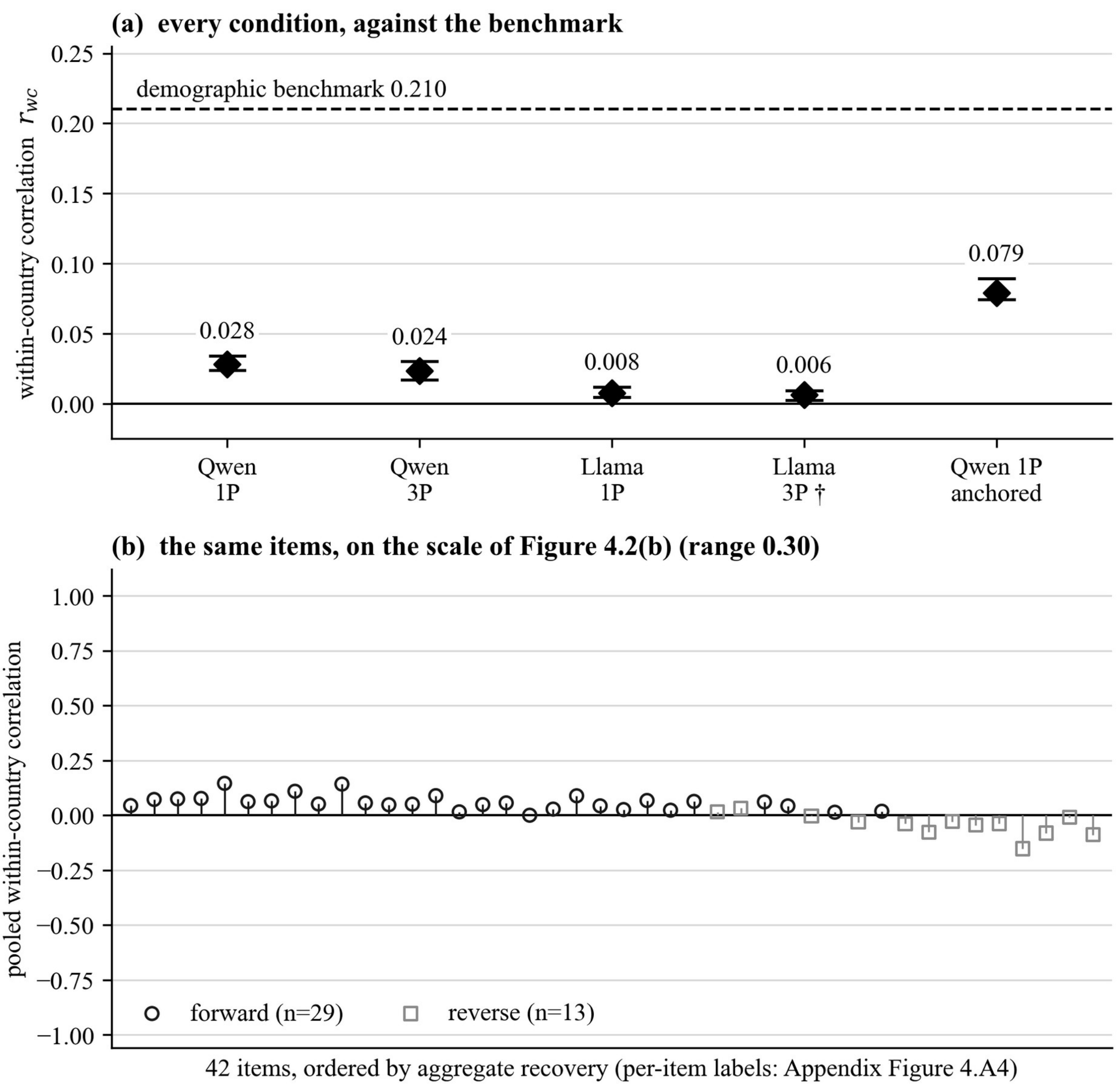

Figure 6: Individual-level recovery by condition and item by item.

*Note:* Panel (a) shows the median within-country correlation in each condition, with a 95% bootstrap interval over country-item cells, against the 0.210 demographic benchmark (Section 3.5). The conditions are those of Figure 2(a), and the dagger marks the Llama 3P arm, which is not read substantively (Section 4.4). The anchored arm is the 22-item batch of Section 3.4. Panel (b) shows the pooled within-country correlation for each of the 42 items, in the item order and on the scale of Figure 2(b). One country-item cell is undefined because the silicon response is constant.

Individual recovery is negligible in every country. The highest country mean is Israel's 0.077, which explains less than 1% of individual variance. Individual recovery varies more by item than by country. Life satisfaction and happiness recover best, both near 0.14, whereas feeling about household income (−0.148) and whether gay men and lesbians should be free to live as they wish (−0.085) are consistently negative. At this level 11 of the 42 items are negative, the same count as at the aggregate level and largely the same items (Appendix Figure A4, and Table A3 by domain). That is the pattern a misread scale direction, operating at both levels, would produce. What is asked matters more than which country is simulated.

Scale anchoring is the only manipulation that raises individual-level recovery (Section 4.2). On the same 22 items, scale anchoring raises median within-country recovery from 0.010 in the numeric arm to 0.079. Recovery stays far below the 0.210 demographic benchmark in both cases.

Model responses are compressed in every substantively read arm. Under Qwen 1P the ratio of model to human standard deviations runs from 0.57 to 0.61 depending on the estimator, far below the prespecified 0.9–1.1 band. Aggregate rank correspondence does not imply distributional recovery.

Aggregate and individual recovery come apart across items and across countries. Across items the aggregate correlations run from −0.82 to +0.82, against a range of 0.30 at the individual level. Across 30 countries, the profile correlation and individual recovery correlate at −0.21. The 90% Fisher-z interval is [−0.49, 0.10], so +0.10 is a one-sided 95% upper bound on how positive the association can be. Of the 435 country pairs, 232 are ordered one way by the profile correlation and the other way by individual recovery (Figure 7), so a researcher choosing between two countries on aggregate evidence would order the pair the wrong way often. Switzerland against Israel is the sharpest case: 0.91 against 0.75 on the profile correlation, and 0.014 against 0.077 within countries. The correlation between the two rankings is zero or negative in the other three conditions. Aggregate recovery therefore cannot be used to choose countries for individual-level work.

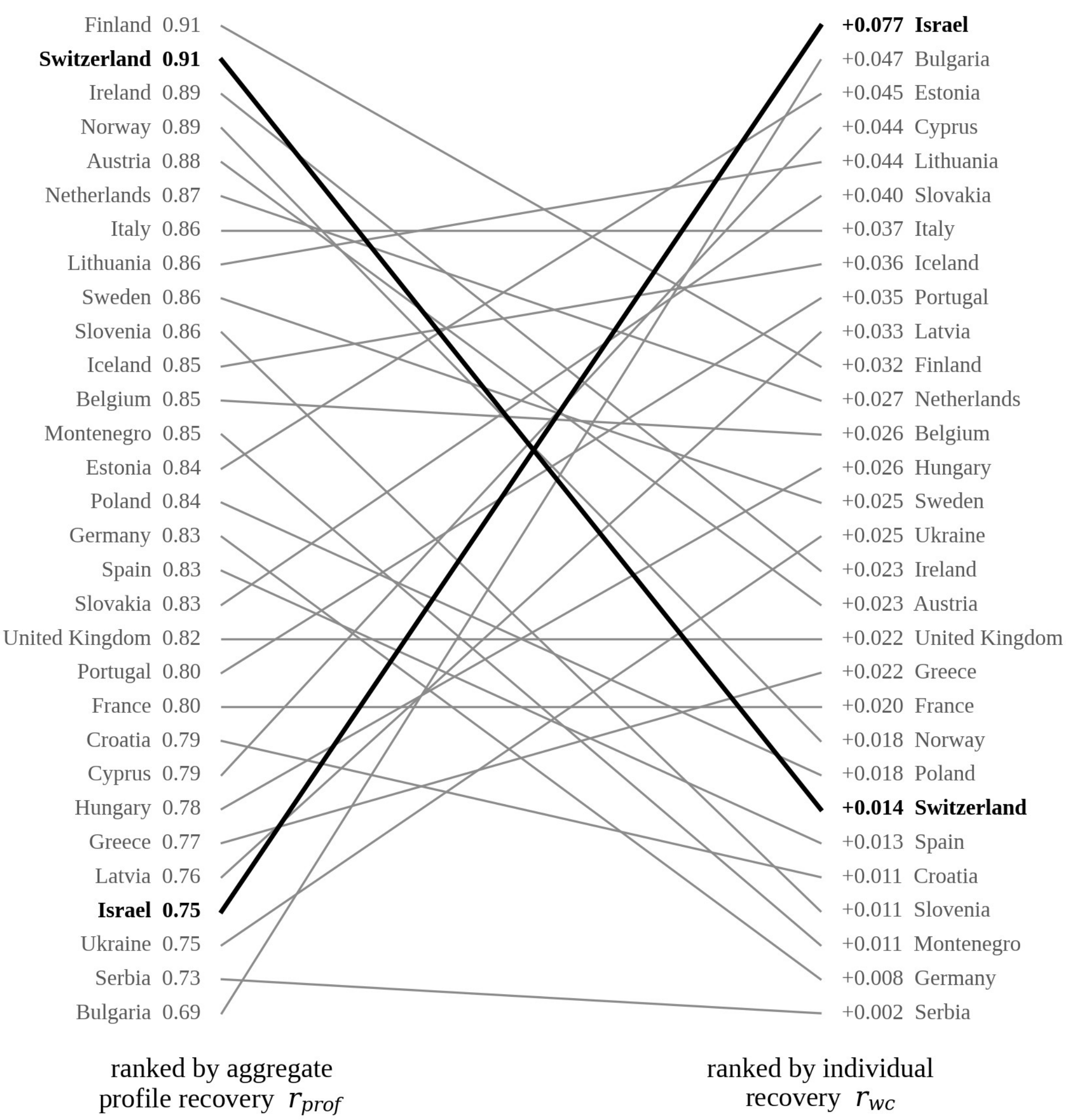

Figure 7: Aggregate against individual country rankings, Qwen 1P.

*Note:* Left: the 30 countries ranked by profile correlation. Right: the same countries ranked by mean $r_{wc}$. Crossing lines are rank reversals, and 232 of the 435 country pairs reverse. Switzerland and Israel are drawn in black.

### 4.4 Robustness checks

The 17 checks leave the conclusions where they stand, with one exception: parse coverage in the Llama third-person arm. Appendix Table A5 gives the quantity each one moves.

The direction coding comes from the ESS codebook and was set before any result was seen. A second model reproduces the resulting item pattern. Item-level recovery correlates across the two models at $r$ = 0.84.

Run-to-run stability sets the floor an effect must clear. The median absolute difference in $r_{bc}$ between the two identical runs is 0.060 per item, and every interpreted raw-correlation effect exceeds that floor. Fisher-z effects are read against the corresponding 0.079 floor. A second replicate on the full battery gives a smaller median, so the floor is kept at 0.060, the conservative choice.

Parse coverage in the Llama third-person arm is the exception. Restricting to items above 80% coverage moves the median $r_{bc}$ from +0.17 to zero. The correlation falls as the threshold tightens, so the usable subset is biased and not cleaner. The poorly covered items are the wide response scales, on which recovery is otherwise strongest. None of the four alternative extraction rules rescues the condition. Re-parsing every rejected response under those rules moves the median $r_{bc}$ in that arm by at most 0.013 (Appendix Table A6), against the 0.060 noise floor. The framing null rests on Qwen alone.

## 5 Discussion and conclusion

### 5.1 The country label acts as a country-level assumption

In silicon sampling, the country label produces moderate aggregate recovery, and richer profiles add no consistent gain (Section 4.2). An LLM that found respondent detail uninformative would show flat effects at both positions. The political and socioeconomic effects are negative, so the model does read that detail. Almost none of the variation the model produces within a country is respondent-specific, and in the replicated arm nearly all of it is random variation between runs (Section 4.3). The label is specific to the country named: replacing it with a wrong name lowers recovery for the respondents' own country.

The model orders countries without matching the distances between them. Calibrated country knowledge would reproduce the spacing and the level of the country means as well as their order. The model compresses cross-national spread, it shifts every country's mean in the same direction, and most of the agreement available at the observed ranking is lost to that displacement (Section 4.1 and Appendix F).

Recovery varies widely among profiles that identify the country equally well, and the sparsest country-labeled profile sits at the top of that range (Appendix I), so identifiability does not explain the pattern. Country may be the single most informative cue for these outcomes, so leaning on it could be sound cue use rather than stereotype activation. A model weighing cues on their validity would not lose accuracy when given more valid ones, yet the socioeconomic block widens cross-national differentiation while lowering the correlation (Section 4.2).

A model can tell countries apart without representing any of them, much as raters agree on national character without matching the traits those populations actually report (Terracciano et al., 2005). The observed correspondence cannot distinguish survey knowledge from online discourse, national stereotypes, or culturally specific scale use. The country label is thus a country-level assumption, not evidence that the model reconstructs the population it names. It answers the cultural-alignment question, which national profiles a model resembles (Durmus et al., 2023; Tao et al., 2024), and leaves the survey-simulation question unanswered, whether conditioned outputs reproduce the target population (Bail, 2024; Bisbee et al., 2024).

### 5.2 Response format explains the inversions in country rankings

The inversions come mainly from the response format, not from the model's country associations. The response-scale experiment of Section 4.2 removes them while holding the model, the respondents, and the items fixed, and it does so in both models. Prompt format is therefore part of the measurement instrument. An ESS respondent hears the full question and sees a labeled show card, while the model receives a bare numeric range. Röttger et al. (2024) reach the same conclusion for the measured political position of models under different question formats. What the numeric arm measures is a compound of the model's country associations and its reading of the scale. Verbal endpoints make the intended scale direction explicit. This is a measurement-equivalence problem of the kind cross-national survey research already recognizes between human populations (Davidov et al., 2014; King et al., 2004), arising here between human and model. A validation must report the response format it used, because the same model and the same items rank countries the other way round under another format. Published benchmarks differ in how they present the answer scale, so part of the disagreement between them may be a difference in format rather than in what the models know.

Most of the variance in item-level recovery follows direction, and what is left has a pattern of its own: institutional-trust items recover relatively well, while several income, health, and safety items invert. A post hoc reading is that the model recovers what is publicly said about a country and not what its residents report about themselves. Institutional trust and subjective well-being have a large descriptive literature and track visible national conditions (Diener et al., 1995; Norris, 2011), while how safe a respondent feels walking home after dark has no comparable public ordering. That reading fits the country-level assumption of Section 5.1, predicts which items a country label can carry, and is consistent with the larger recovery where true cross-national differences are larger (Section 4.1).

In these data, high pooled and profile correlations coexist with the inversions, because both combine well-recovered and inverted items, so a screening rule based on either would pass the battery unchanged. Item-level diagnostics are therefore necessary before pooled or profile metrics are used.

### 5.3 Silicon sampling can support country ranking, not individual or distributional inference

Silicon sampling may support exploratory country-ranking comparison on topics that recover well in an external benchmark and where no survey data from comparable countries are available. It does not support individual prediction, within-country heterogeneity estimation, or distributional analysis: matching country means can coexist with compressed variance, unstable coefficients, and incorrect dependence structure (Bisbee et al., 2024; Williams et al., 2026; Xie et al., 2026). Aggregate recovery is evidence about the ordering of country means alone, and it does not establish representativeness. Validation should match the quantity a study intends to estimate and the design it uses, since response coding, backstory wording, model choice, and sampling settings all change synthetic distributions (Bisbee et al., 2024; Dominguez-Olmedo et al., 2024; Röttger et al., 2024).

Aggregate and individual recovery rest on different things. Relationships among group means need not reproduce relationships among individuals (Firebaugh, 1978; King, 1997; Robinson, 1950). The aggregate correspondence comes mainly from the country label rather than from the individuals, so evidence for aggregate recovery supports a different claim from evidence for individual recovery. Recent evaluations point in the same direction, as LLM-generated data can match marginal statistics while failing multivariate and subgroup structure (Li et al., 2025; Williams et al., 2026). LLM stand-ins reproduce group-level discourse while flattening the within-group variation on which individual recovery depends (Kozlowski & Evans, 2025).

Closing the gap needs more information than a typical survey carries. Higher individual recovery appears only when models receive interview- or annotator-specific data far richer than demographic profiles (Orlikowski et al., 2025; Park et al., 2024), and sociodemographic prompting on its own does not produce it (Gao et al., 2025; Sun et al., 2025).

Whether the recovery from silicon sampling is adequate can depend on the alternative. Both comparators in Section 4.1 are built from survey data in other countries, which a researcher running a cross-national validation already holds. The question is therefore whether such a researcher should use the model to extend those data, and on these items, at this sample size, the answer is no. Pataranutaporn et al. (2025) reach the same conclusion for life satisfaction, where ordinary regression models outperform four LLMs across 64 countries.

### 5.4 Contributions, limitations, and future directions

The empirical contribution is evidence that the country label functions as a country-level assumption. Item rankings, individual recovery, and the label's incremental contribution are measured together on the same respondents, items, and prompts. The methodological contribution is to make those readings testable. Each non-base block is observed at both of its positions, so no ordering of additions can be chosen after the fact. A swapped-label arm tests whether the label is specific to the country named. A 2 × 2 design tests whether response format explains the item inversions. Recovery is reported item by item against a measured noise floor, because a pooled correlation would have hidden the inverted items. The practical contribution is a standard of adequacy. Recovery is judged against comparators that use no LLM, built from survey data a researcher already holds.

The design sets the limits of these conclusions. English prompts and generic party coding omit country-specific language and party cues. One European survey program and two open-weight 7–8-billion-parameter models limit generalization to non-European populations, other survey designs, and larger or proprietary models. The no-country arm still carries country information in the composition of harmonized categories, where a classifier recovers the country 28.7% of the time against a 3.3% baseline. The no-country arm's median recovery of 0.40 on forward-coded items therefore bounds what composition achieves without a country name. No experiment in this design separates calibrated country knowledge from stereotype activation. The Llama third-person arm lost 23% of scheduled responses because the model failed to answer in the required format, which bounds what that arm can show (Appendix L).

A factorial design can vary prompt language, country-specific party cues, verbal endpoints, and profile content independently. Validation should extend to non-European surveys and to larger or proprietary models. Where a design cannot remove the country information carried by sample composition, reporting it makes the residual visible. Separating calibrated country knowledge from stereotype activation needs a manipulation that varies one without the other.

Reporting unparsable outputs by item and country, rather than as one coverage figure, shows where the loss falls. Each use of silicon sampling should be judged against the quantity it is meant to estimate.

## Declarations

**Data availability.** The replication materials, comprising the analysis-ready model summaries, the analysis and figure code, and the raw model responses paired with their human benchmark values, are available at https://github.com/chuyao-wang/silicon-replication. The European Social Survey Round 11 data are publicly available from the European Social Survey Data Portal; the edition used is cited in the references.

**Ethics.** [Confirm: The study analyzes publicly available survey data and model-generated responses and did not involve the collection of new data from human participants.]

**Funding.** [Insert the funding statement, or state that the research received no specific grant from any funding agency.]

**Acknowledgments.** The author thanks the London School of Economics and Political Science for access to the Fabian high-performance computing cluster, which provided the GPU resources used in this study.

**Declaration of competing interests.** The author declares no competing interests.

## Appendix

The appendix has 13 sections (A to M), with figures and tables numbered consecutively from A1: Table A1 in A, Table A2 in C, Figures A1–A2 in E, Figure A3 in G, Tables A3–A4 and Figure A4 in J, Table A5 in K, and Table A6 in M.

## A Survey items

Table A1 documents the 42 ESS items, their abbreviated question wording, their numeric response scales, and the domain groupings used in the item- and domain-level analyses.

Table A1: The 42 ESS Round 11 survey items.

| Variable | Question text | Scale | Domain |
|---|---|---|---|
| health | Subjective general health | 1-5 | Health |
| imbgeco | Immigration bad or good for country's economy | 0-10 | Immigration Attitudes |
| imdfetn | Allow many/few immigrants of different race/ethnic group from majority | 1-4 | Immigration Attitudes |
| impcntr | Allow many/few immigrants from poorer countries outside Europe | 1-4 | Immigration Attitudes |
| imsmetn | Allow many/few immigrants of same race/ethnic group as majority | 1-4 | Immigration Attitudes |
| imueclt | Country's cultural life undermined or enriched by immigrants | 0-10 | Immigration Attitudes |
| imwbcnt | Immigrants make country worse or better place to live | 0-10 | Immigration Attitudes |
| hincfel | Feeling about household's income nowadays | 1-4 | Income |
| trstep | Trust in the European Parliament | 0-10 | Institutional Trust |
| trstlgl | Trust in the legal system | 0-10 | Institutional Trust |
| trstplc | Trust in the police | 0-10 | Institutional Trust |
| trstplt | Trust in politicians | 0-10 | Institutional Trust |
| trstprl | Trust in country's parliament | 0-10 | Institutional Trust |
| trstprt | Trust in political parties | 0-10 | Institutional Trust |
| trstun | Trust in the United Nations | 0-10 | Institutional Trust |
| atchctr | How emotionally attached to country | 0-10 | National Attachment |
| atcherp | How emotionally attached to Europe | 0-10 | National Attachment |
| polintr | How interested in politics | 1-4 | Political Attitudes |
| psppipla | Political system allows people to have influence on politics | 1-5 | Political Attitudes |
| psppsgva | Political system allows people to have a say in what government does | 1-5 | Political Attitudes |

| Variable | Question text | Scale | Domain |
|---|---|---|---|
| stfdem | How satisfied with the way democracy works in country | 0-10 | Political Attitudes |
| vote | Voted last national election | 1-3 | Political Attitudes |
| actrolga | Able to take active role in political group | 1-5 | Political Efficacy |
| cptppola | Confident in own ability to participate in politics | 1-5 | Political Efficacy |
| rlgatnd | How often attend religious services apart from special occasions | 1-7 | Religion |
| aesfdrk | Feeling of safety of walking alone in local area after dark | 1-4 | Safety |
| inprdsc | How many people with whom you can discuss intimate and personal matters | 0-6 | Social Trust |
| pplfair | Most people try to take advantage of you, or try to be fair | 0-10 | Social Trust |
| pplhlp | Most of the time people are helpful or mostly looking out for themselves | 0-10 | Social Trust |
| ppltrst | Most people can be trusted, or you can’t be too careful | 0-10 | Social Trust |
| sclmeet | How often socially meet with friends, relatives or colleagues | 1-7 | Social Trust |
| euftf | European unification go further or gone too far | 0-10 | Social Values |
| freehms | Gay men and lesbians free to live life as they wish | 1-5 | Social Values |
| gincdif | Government should reduce differences in income levels | 1-5 | Social Values |
| hmsacld | Gay male and lesbian couples should have right to adopt children | 1-5 | Social Values |
| hmsfmlsh | Ashamed if close family member is gay or lesbian | 1-5 | Social Values |
| happy | How happy are you | 0-10 | Subjective Well-being |
| stfeco | How satisfied with present state of economy in country | 0-10 | Subjective Well-being |
| stfedu | State of education in country nowadays | 0-10 | Subjective Well-being |
| stfgov | How satisfied with the national government | 0-10 | Subjective Well-being |
| stfhlth | State of health services in country nowadays | 0-10 | Subjective Well-being |
| stflife | How satisfied with life as a whole | 0-10 | Subjective Well-being |

*Note:* Scale is the numeric range supplied to the model, with no verbal endpoints except in the three social trust questions. Domain is the grouping used in the domain-level analyses.

## B Example prompt

The abbreviated example illustrates the first-person framing. It is reproduced as the model received it, including the scale line, which states the numeric range and names no endpoint. Only the respondent backstory is shortened for display.

Adopt the following persona and respond as if you were this person.

I live in Germany. I am female. I am 52 years old. I am married. I live in a big city. My highest level of education is upper secondary, with 12 years of full-time education. My main activity is paid work. My household income is in the 6th decile. [...]

Question: Using this card, generally speaking, would you say that most people can be trusted, or that you can't be too careful in dealing with people? Please tell me on a score of 0 to 10, where 0 means you can't be too careful and 10 means that most people can be trusted.

Scale: 0-10

Respond with ONLY a single number within the scale range. No explanation.

## C Backstory leakage

Table A2 shows how removing leaked outcome information changes the seven affected items. Its final column distinguishes direct echoes from attenuated inverse mappings.

Table A2: Backstory leakage for the seven overlapping variables.

| **Variable** | **Pooled individual r (leaked)** | **Pooled individual r (clean)** | **Difference** | **Interpretation** |
|---|---|---|---|---|
| inprdsc | 0.854 | 0.038 | −0.817 | Echo removed |
| sclmeet | 0.291 | 0.035 | −0.256 | Echo removed |
| vote | 0.142 | −0.037 | −0.179 | Echo removed |
| hincfel | −0.184 | −0.164 | +0.020 | Attenuated |
| health | −0.327 | −0.093 | +0.234 | Attenuated |
| rlgatnd | −0.371 | 0.014 | +0.385 | Attenuated |
| polintr | −0.265 | −0.092 | +0.173 | Attenuated |

*Note:* “Echo removed” indicates variables where the leaked backstory directly stated the answer. “Attenuated” indicates variables where leakage was associated with more negative correlations, consistent with inverse scale mapping. These correlations moved toward zero once the leaked variable was removed. Across all 42 items, the leaked backstory raised the mean pooled individual correlation from 0.044 (clean) to 0.070. The leaked column derives from the 27-variable-backstory run (Qwen, first-person, temperature 0.7, approximately 400 respondents per country) and the clean column from the 20-variable run (approximately 500 respondents per country). Both predate the final 685-respondent round. The comparison is retained as the design rationale for removing the seven overlapping variables.

## D Evaluation metrics

Section 3.5 defines $r_{bc}$ and $r_{wc}$, and Section 4.1 the profile correlation. One further measure appears only in the appendix: an uncentered pooled individual correlation, used in Table A2 and in the specification check in Table A5. Because it retains between-country covariance, it is not a measure of individual-level prediction.

## E Per-item aggregate recovery

Figure A1 reports the full per-item aggregate recovery that Figure 2(b) shows unlabeled.

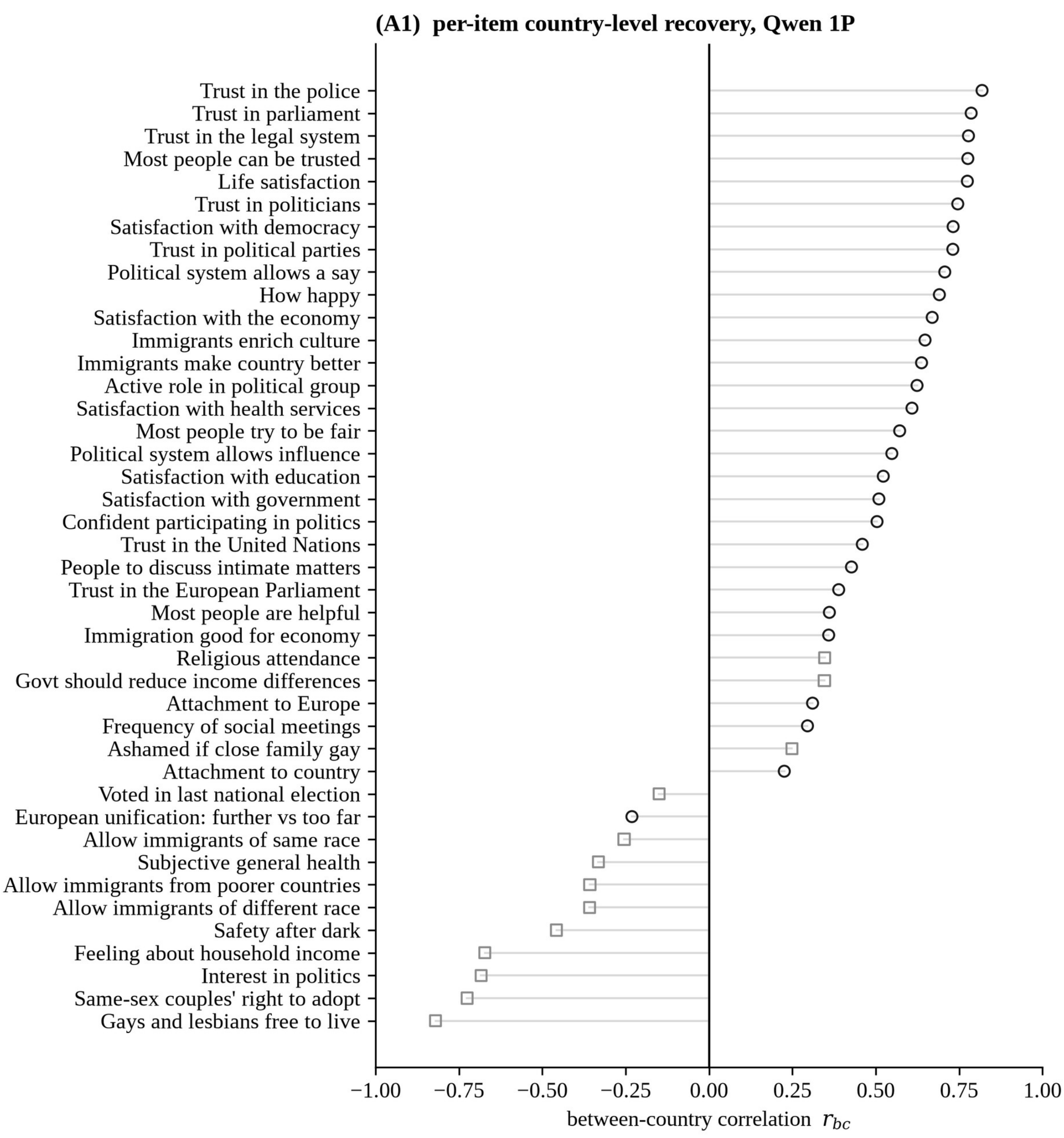

Figure A1: Per-item aggregate recovery, Qwen 1P.

*Note:* $r_{bc}$ for all 42 ESS items, ranked from most positive to most negative. Circles are forward-coded items and squares reverse-coded, as in Figure 2(b).

Item-level recovery can reflect either an accurate country ordering or a systematic inversion, while compressing cross-national differences in both cases. Figure A2 shows both directly, and that compression is what the concordance coefficient below puts a number on.

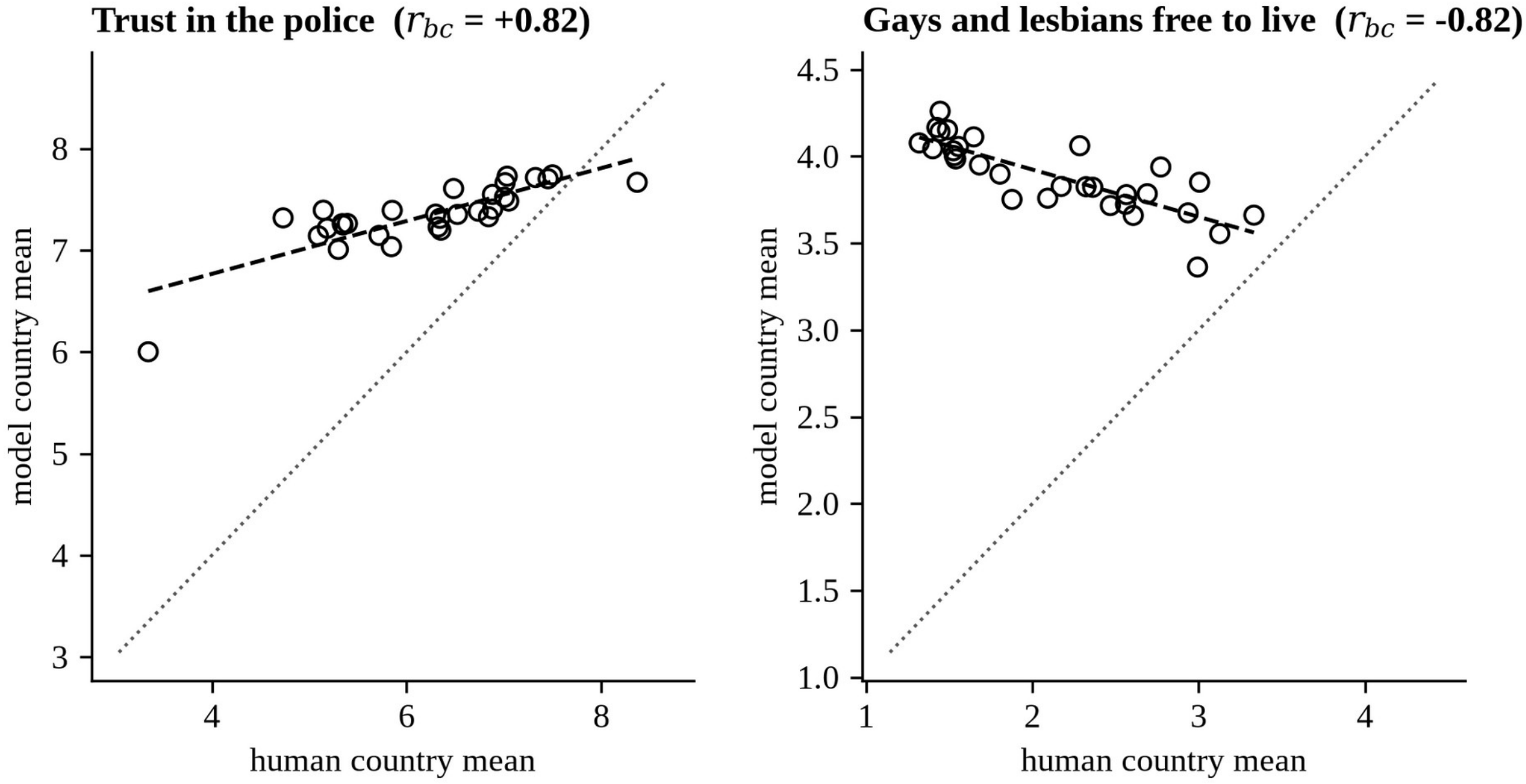

Figure A2: Country-level scatter for two contrasting items, Qwen 1P.

*Note:* Left: trust in police ($r_{bc}$ = 0.82), with the country ordering largely recovered. Right: attitudes toward homosexuality ($r_{bc}$ = −0.82), with the ordering inverted. Dashed lines show least-squares fits, and dotted lines show the 45-degree reference for perfect recovery.

## F Concordance

Pearson's r is invariant to location and scale, so a model whose country means are a compressed, shifted copy of the human ones still scores 1. Lin's concordance coefficient rewards agreement with the 45-degree line and factors as $r_{bc} \times C_b$, where $C_b$ is an accuracy term bounded by 1. Under Qwen 1P, the median accuracy term is $C_b$ = 0.132, so 87% of the agreement attainable at the observed ranking is lost to bias and compression. The separate medians are $r_{bc}$ = 0.443 and concordance = 0.036. The median absolute bias is 1.13 scale points and the median between-country dispersion ratio 0.28. Of the 42 items, 11 have negative concordance, which is worse than no agreement at all.

## G The country label contrast

Figure A3 reports the country label contrast item by item, the per-item form of the block effect in Figure 4.

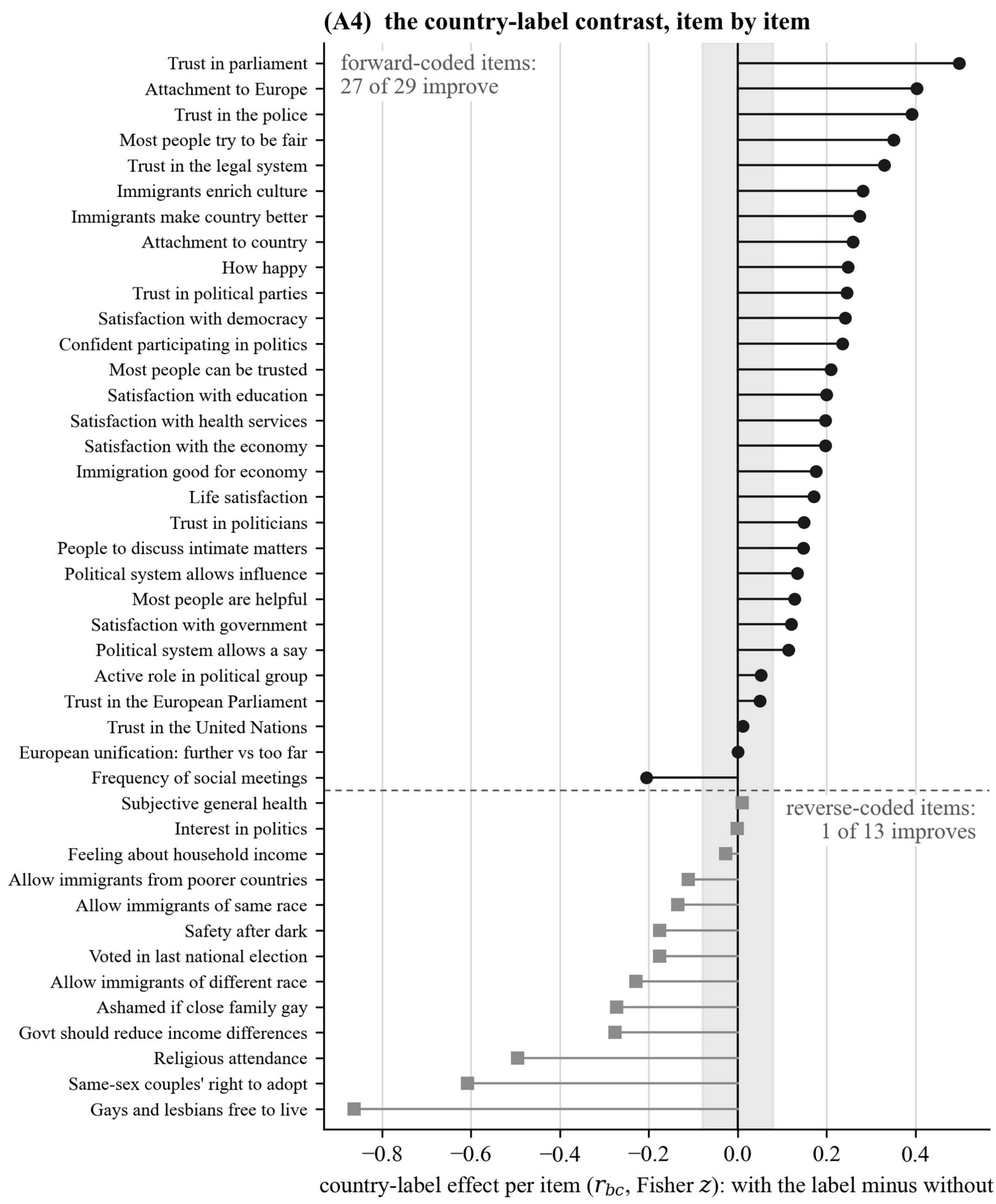

Figure A3: The country label contrast, item by item, Qwen 1P.

*Note:* The change in $r_{bc}$ (Fisher z) for each of the 42 items when the country sentence is present in an otherwise identical full profile, the contrast Figure 4 summarizes as one bar. Items are ranked within wording direction. The shaded band is the Fisher-z noise floor (±0.079).

## H Block contrasts

The two positions of Section 3.4 run on the same respondents throughout, so every comparison is paired at the item level. Among the three original removal contrasts, only the country label produces an effect that is both significant after Holm correction and larger than the 0.060 noise floor: +0.135 on forward items and −0.176 on reverse items. The region code (+0.020 forward) and the political variables (+0.042 reverse) point the same way but fall inside the noise floor, so they are reported as negligible. Across the remaining blocks the largest positive forward effect at either position is +0.09 (Fisher z), the union and internet block, and the single-variable arms attribute it to internet use. Per-block and per-item effects are provided with the Figure 4 data.

## I Country identifiability

A multinomial classifier was trained to recover the country from the rendered backstory text of each of the 10 variants, with a majority-class baseline of 3.33%. Eight of them reach exactly 100%, because each renders the country sentence, so identifiability is saturated and cannot produce a dose-response. At that identical level, median $r_{bc}$ still ranges from 0.287 for the superseded 14-variable political variant to 0.523 for the four-variable minimal profile. The three-variable base reaches 7.23%, and the full profile without its country sentence 28.72%, replicated at 28.99% in an independent run. Its strongest n-grams are fragments of harmonized education, marital status, and main activity, and no rendered backstory names a country, so the residual identifiability is compositional.

## J Individual-level diagnostics

As a conservative demographic benchmark, human responses for each item are regressed on age, gender, education years, household income decile, and left-right self-placement, with country fixed effects. The median within-country multiple correlation is 0.210 (interquartile range 0.15–0.26, maximum 0.47) and exceeds 0.20 for 23 of 42 items. Mean $r_{wc}$ of 0.027 therefore recovers only a small fraction of the available demographic signal. The most predictable human items are among the model's weakest: household income feeling (benchmark 0.47, model −0.148), subjective health (0.42, −0.078), and political interest (0.32, −0.079).

Table A3 breaks individual-level recovery down by domain. Political efficacy, institutional trust, and subjective well-being recover best, while income, health, safety, and social values are negative. The largest domain mean in any condition is 0.125, against the 0.210 demographic benchmark in Section 3.5, so no domain reaches even two-thirds of what five demographic predictors recover linearly.

Table A3: Mean $r_{wc}$ by domain and model-prompt condition.

| Domain | Qwen 1P | Qwen 3P | Llama 1P | Llama 3P |
|---|---|---|---|---|
| National Attachment | 0.036 | 0.103 | 0.021 | 0.006 |
| Political Efficacy | 0.088 | 0.125 | 0.027 | 0.070 |
| Health | −0.078 | −0.056 | −0.046 | −0.021 |
| Immigration Attitudes | 0.008 | 0.000 | −0.005 | −0.007 |
| Income | −0.148 | −0.133 | −0.052 | −0.022 |
| Institutional Trust | 0.069 | 0.068 | 0.017 | 0.008 |
| Political Attitudes | 0.013 | 0.009 | 0.008 | 0.014 |
| Religion | 0.012 | −0.062 | −0.008 | 0.018 |
| Safety | −0.037 | −0.050 | −0.002 | −0.002 |
| Social Trust | 0.043 | 0.024 | 0.021 | 0.013 |
| Social Values | −0.011 | −0.016 | −0.008 | −0.006 |
| Subjective Well-being | 0.066 | 0.082 | 0.034 | 0.000 |

*Note:* Each cell is the mean within-country correlation over the items in that domain. The demographic benchmark for comparison is 0.210 (Section 3.5).

Of the 42 items, 11 show negative $r_{wc}$ under Qwen 1P, meaning the model's predictions are inversely related to human responses. These cases are concentrated in reverse-coded items, for which substantive inversion cannot be separated from scale-direction misinterpretation (Figure A4).

Figure A4 reports the full per-item individual-level recovery.

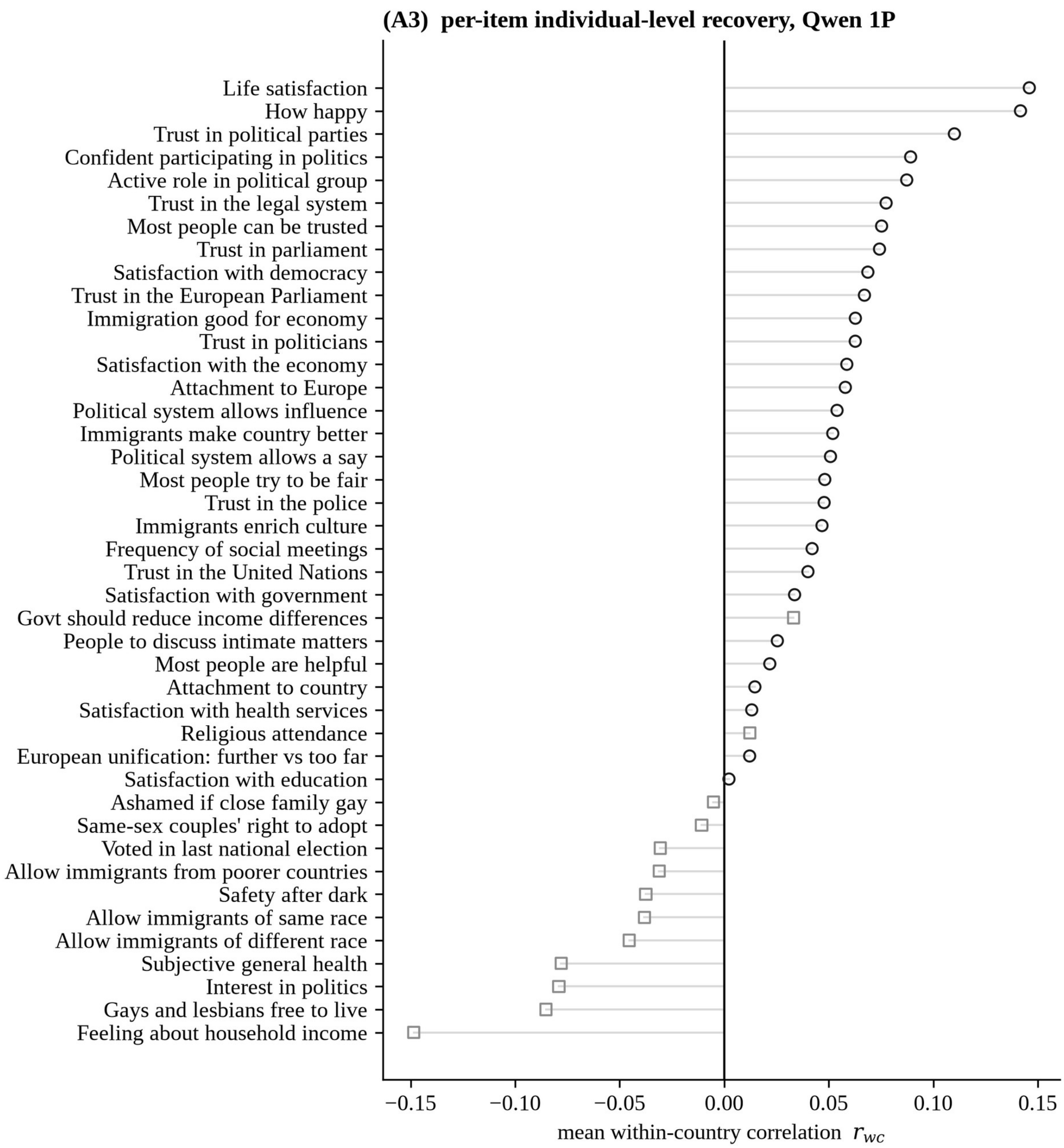

Figure A4: Per-item individual-level recovery, Qwen 1P.

*Note:* Mean within-country $r_{wc}$ for all 42 ESS items, ranked from most positive to most negative. The individual-level counterpart of Figure A1, ranked on its own values. Figure 6(b) shows the pooled-within counterpart in the aggregate item order.

Table A4 reports the prespecified model and prompt contrasts on the per-item mean $r_{wc}$. The Qwen prompt-framing difference is small and far from significance.

Table A4: Paired t-tests for model and prompt comparisons on mean $r_{wc}$.

| Comparison | $r_{wc}$ (A) | $r_{wc}$ (B) | Diff | Std. error | t | p |
|---|---|---|---|---|---|---|
| Qwen: 1P vs. 3P | 0.027 | 0.029 | −0.002 | 0.006 | −0.28 | 0.785 |
| Llama: 1P vs. 3P | 0.009 | 0.006 | +0.004 | 0.004 | 0.99 | 0.326 |
| 1P: Qwen vs. Llama | 0.027 | 0.009 | +0.018 | 0.006 | 2.81 | 0.007 |
| 3P: Qwen vs. Llama | 0.029 | 0.006 | +0.023 | 0.010 | 2.28 | 0.028 |

*Note: N = 42 items for every test. After Bonferroni correction for four tests (α = 0.0125), the first-person Qwen-Llama contrast reaches the adjusted threshold. The other three do not. Reported p-values are unadjusted. The Llama 1P versus 3P contrast is not read substantively, because the third-person arm's parse coverage cannot identify a framing effect (Section 4.4).*

## K Aggregate robustness checks

Table A5 lists all 17 checks with the quantity each moves. One changes a conclusion, and Section 4.4 states it. Three others turn on the choice of estimator. The within-country dispersion ratio has three defensible aggregators (mean of ratios 0.61, median 0.57, ratio of means 0.59) and all three fail the prespecified 0.9–1.1 band, so the choice is named wherever the quantity appears and no conclusion rests on it. Design weighting correlates the weighted and unweighted per-item coefficients at 0.999 while moving a single item by up to 0.103, so the item-level maximum is reported. The direction classification was frozen from the ESS codebook before any result was seen, and the direction $R^2$ rises from 0.658 to 0.746 when vote and the two pre-declared uninterpretable items are dropped, so the association strengthens under this exclusion.

Table A5: The 17 robustness checks.

| Check | What it varies | Result | Any conclusion change? |
|---|---|---|---|
| Design weighting of the human benchmark | Human country means computed unweighted instead of with pspwght | The two coefficient vectors correlate at 0.9987; the largest item-level change is 0.103 (health) | No. The median and the direction decomposition are unaffected. |
| Direction classification of the 42 items | The frozen classification against the four disputed codings | $R^2$ = 0.6576 on all 42 items, 0.6531 excluding vote, 0.7455 further excluding the two screened items | No. The effect strengthens as the least interpretable items are removed. |

| Check | What it varies | Result | Any conclusion change? |
|---|---|---|---|
| Items whose silicon dispersion is near zero | A reliability screen on the silicon side | Two items fall below it: hmsfmlsh (reliability 0.25) and rlgatnd (0.44); inprdsc (0.615) is retained | No. Both are prespecified uninterpretable and the primary anchored specification excludes them. |
| Item set for the anchoring experiment | Three prespecified reverse-item sets | All 12 in batch: +0.4912 [0.1515, 0.8809], 11 of 12 up; primary 10: +0.4230 [0.1259, 0.7928], 10 of 10 up; strictest 8: +0.1636 [0.0507, 0.2671], 8 of 8 up | No for the sign and the interval. Yes for the gap verdict: eliminated on the primary set, narrowed with a residual on the strictest. |
| Choice of forward control items | 10 leave-one-out control sets | The triple difference ranges from 0.403 to 0.455 across the 10 sets, against the primary 0.423; the largest shift is 0.032 | No. |
| Model for the anchoring experiment | Llama 1P on the same 22 items and the same respondents | Triple difference +0.4488 [0.1551, 0.7374] on the primary set of 10, 8 of 10 up; +0.5446 [0.2324, 0.8823] on all 12 in batch and +0.3060 [0.0381, 0.5719] on the strictest 8 | No. Under Llama, the forward-reverse gap closes from +0.3544 to −0.0944, a slight overshoot. Under Qwen, it lands inside the noise band. |
| Run-to-run stability of the generation | Two identical unseeded runs | Median absolute $r_{bc}$ difference 0.0602 per item, maximum 0.1854, no sign reversals | No, and it closes a stated limitation: raw-correlation contrasts are read against the 0.060 band and Fisher-z contrasts against the 0.079 band. |
| Deflator for the macro-covariate comparator | Market dollars against purchasing parity | Median $r_{bc}$ 0.6177 under dollars, 0.6317 under parity; normalized error 0.0522 and 0.0527 | No for the comparison with the model. Yes for one claim, therefore not made: the covariate-alone ranking is deflator-dependent. |
| Israel as a bloc of one in the comparator | Israel excluded from every arm | The comparator's median $r_{bc}$ moves from 0.6177 to 0.6741; the gap to the model from 0.174 to 0.177, with every arm recomputed on the 29-country sample | No. |
| Influential countries in the comparator | Leave-one-out over the 30 countries | Excluding Ireland moves the reported value by 0.001; the high-leverage country is Ukraine (mean hat 0.495) | No. |
| Coverage in the Llama third-person arm | Restricting to items above 80% coverage | Median $r_{bc}$ falls from +0.1729 on all 42 items to −0.0009 on the 19 items above the threshold | Yes. The arm cannot identify a framing effect; the framing null rests on Qwen alone. |
| Aggregator for the dispersion ratios | Mean of ratios, median, and ratio of means | Within-country 0.6052, 0.5727, 0.5859; between-country 0.2821 median, 0.2952 mean | No. All three within-country estimators fail the prespecified 0.9–1.1 band. |

| Check | What it varies | Result | Any conclusion change? |
| --- | --- | --- | --- |
| Estimator for the offset reduction | Four defensible estimators | Ratio of medians 64.2%, ratio of means 62.9%, median of per-item reductions 59.9%, mean of per-item reductions 53.7% | The offset reduction is the fall in an item's normalized error when its constant level offset is removed. The conclusion is unchanged under all four estimators. |
| Specification for the cross-level agreement | Item coding (as coded, forward-only, sign-aligned) and the individual-level series (uncentered pooled individual, pooled-within) | Against the uncentered pooled individual series: Pearson 0.9368 as coded, 0.7745 forward-only, and 0.8355 sign-aligned; against the pooled-within series: 0.8540, 0.4899, and 0.6326 | No for the existence of the agreement. Yes for its description: about a third of the shared variance on the uncentered series, and about two-thirds on the pooled-within series, reflects the same misread scale direction. |
| Frame for country-level quantities | Full frame against the human-model intersection | Median $r_{bc}$ moves from 0.4433 to 0.4425; single items move by up to 0.109 (health) | No for the median. The frame is fixed in one place and stated. |
| Stability across rounds | The March round against the final round | Median absolute item-level $r_{bc}$ change 0.0415, against a same-round replicate median of 0.060 | No. |
| Human-side missingness rule | Correcting the missing-code set for the coarse scales | Median $r_{bc}$ 0.418 to 0.431, assessed retrospectively on the earlier release that motivated the correction; items above 0.70 from 8 to 9; inverted items from 12 to 11; within-country dispersion ratio 0.553 to 0.605. The final release applies the corrected rule at generation time; the reported values (median 0.443) are computed under it | No for any conclusion. Corrected several reported figures. |

*Note:* Each row names what the check varies, what it moves, and whether any conclusion changes. The one Yes that changes a conclusion is parse coverage in the Llama third-person arm (Section 4.4). Bracketed pairs are 95% bootstrap intervals over items; ranges across specifications are written out in words.

Computing each country mean on only the responses with both a valid human answer and a parseable model answer moves per-item between-country correlations by at most 0.109. Two further checks ran on the earlier 500-respondent round: excluding the voting item's 'not eligible' category, and replacing the human subsample with the full ESS sample. Both left the per-item correlations at $r$ = 0.99 and preserved every negative correlation, and the cross-round stability check in Table A5 extends them to the final 685-respondent round.

## L Parse failures

Of the 863,100 responses scheduled in the Llama third-person arm, 77.28% parsed. Across all scheduled responses, 5.75% contain an out-of-scale numeric value whose absolute value is at most 1,000, 8.25% contain a numeric value whose absolute value exceeds 1,000, and 8.71% contain no digit. No single source accounts for most of these failures. Truncation: failed outputs have a median length of 61 characters, the longest 317, and none falls within 2% of the maximum output length. Overshoot: 93.5% of out-of-range values lie above the top of the scale, but the median excess is 39 scale points, 6.7 times the scale width. Scale confusion: the out-of-range values take 2,963 distinct values instead of clustering on values valid elsewhere in the battery. Backstory echo: 2.33% of rejected values match one of that respondent's own backstory numbers, against 0.94% under a shuffled null. Reading the verbatim outputs shows what does occur: code blocks, enumeration of the scale in place of a choice within it, continuation of the prompt, and a junk number followed by a valid answer. The first-person control parsed 99.99% with 64 rejections, almost all fractional midpoints, so the difference is the framing, not the model.

## M Sensitivity to the extraction rule

Every correlation reported here is computed on responses that survived a rule that takes the first number in the response text. That rule discards answers the model gave correctly, such as a junk number followed by a valid one, so all 11.3 million raw responses were re-parsed and every rejected response in the 15 arms available at the time of the check was re-scored under four alternative rules: the last number in the text, the number following an answer marker, the only in-range value where exactly one exists, and the first in-range value anywhere. The fourth is the loosest and is reported as an upper bound. In the Llama third-person arm, which holds almost all the rejections, 22.6% of them carry an in-range value somewhere in the text; no arm exceeds 38%. No rule moves a median between-country correlation by more than 0.013, and the primary Qwen first-person arm does not move at all under any rule (+0.4433 throughout). The largest movement in any single item anywhere is 0.092, in the demographics-only arm, whose median is near zero under every rule. The same pass recomputes each released country mean from the raw responses as a check on the accumulator. The worst disagreement is $1.8 \times 10^{-15}$. Table A6 reports all 15 arms.

Table A6: Sensitivity of aggregate recovery to the extraction rule.

| Arm | Responses rejected | Of those, carry an in-range value | Median $r_{bc}$, primary rule | Range across the four rules | Largest shift in any single item |
|---|---|---|---|---|---|
| Qwen 1P | 619 (0.07%) | 0 (0.0%) | +0.4433 | +0.4433 to +0.4433 | 0.0000 |
| Qwen 3P | 71 (0.01%) | 5 (7.0%) | +0.4426 | +0.4426 to +0.4426 | 0.0007 |
| Llama 1P | 64 (0.01%) | 24 (37.5%) | +0.3486 | +0.3486 to +0.3486 | 0.0072 |
| Llama 3P | 196,067 (22.72%) | 44,374 (22.6%) | +0.1729 | +0.1603 to +0.1798 | 0.0596 |
| Demographics only | 6,333 (0.73%) | 419 (6.6%) | −0.0286 | −0.0363 to −0.0286 | 0.0919 |
| + country | 10,729 (1.24%) | 165 (1.5%) | +0.5232 | +0.5232 to +0.5239 | 0.0067 |
| Superseded socioeconomic arm (13 variables) | 1,716 (0.20%) | 6 (0.3%) | +0.3967 | +0.3967 to +0.3967 | 0.0005 |
| Superseded political arm (14 variables) | 1,307 (0.15%) | 2 (0.2%) | +0.2870 | +0.2870 to +0.2870 | 0.0003 |
| No country | 409 (0.05%) | 1 (0.2%) | +0.3441 | +0.3441 to +0.3441 | 0.0007 |
| No region | 648 (0.08%) | 1 (0.2%) | +0.4157 | +0.4157 to +0.4157 | 0.0000 |
| No political | 808 (0.09%) | 0 (0.0%) | +0.5231 | +0.5231 to +0.5231 | 0.0000 |
| Qwen 1P anchored | 53 (0.01%) | 0 (0.0%) | +0.6501 | +0.6501 to +0.6501 | 0.0000 |
| No country, anchored | 75 (0.02%) | 0 (0.0%) | +0.3891 | +0.3891 to +0.3891 | 0.0000 |
| No region, anchored | 46 (0.01%) | 0 (0.0%) | +0.6369 | +0.6369 to +0.6369 | 0.0000 |
| Repeat run (Qwen 1P, 22-item batch) | 67 (0.01%) | 1 (1.5%) | +0.3249 | +0.3249 to +0.3249 | 0.0003 |

*Note:* The four alternative extraction rules are those described above. The primary rule takes the first number in the response text. Arm medians in this table are levels, not paired contrasts; the block effects of Section 4.2 are computed item by item on matched arm pairs, so differences between medians here do not estimate them.